\documentclass[twoside,onecolumn,11pt]{article} 
\usepackage[a4paper,margin=2.5cm]{geometry}
\usepackage[version=3]{mhchem}
\usepackage{balance}
\usepackage{mathtools,amssymb,amsthm} 
\usepackage{sectsty}
\usepackage{graphicx} 
\usepackage{lastpage}
\usepackage[format=plain,justification=justified,singlelinecheck=false,font={stretch=1.125,small,sf},labelfont=bf,labelsep=space]{caption}
\usepackage{float}
\usepackage{fnpos}
\usepackage{cancel}
\usepackage[english]{babel}
\addto{\captionsenglish}{%
  
}
\usepackage[bibstyle=apa,
natbib=true, 
citestyle=authoryear, 
maxnames=1, 
maxcitenames=2,
maxbibnames=9, 
minbibnames=1,  
hyperref=true,  
sorting=nty,
url=true,
uniquename=false,
backend=biber]{biblatex}
\AtEveryBibitem{\clearfield{month}} 
\DeclareFieldFormat{labeldate}{\printdate}
\AtEveryCitekey{\clearfield{month}}
\usepackage{xurl} 
\DeclareFieldFormat{url}{URL: \normalfont\href{#1}{#1}}
\DeclareFieldFormat{doi}{DOI: \normalfont\url{https://doi.org/#1}}
\DeclareFieldFormat{urldate}{}

\usepackage{array}
\usepackage{droidsans}
\usepackage{charter}
\usepackage[T1]{fontenc}
\usepackage[dvipsnames,table,xcdraw]{xcolor}
\usepackage{titlesec}
\usepackage{hyperref}
\usepackage{lineno} 

\usepackage[capitalise]{cleveref}
\crefname{figure}{Figure}{Figures} 
\crefname{table}{Table}{Tables} 
\crefname{equation}{Equation}{Equations} 
\crefname{algocf}{Algorithm}{Algorithms} 

\usepackage[hybrid]{markdown}
\usepackage{tabularx}
\usepackage{tabu}
\usepackage{longtable}
\usepackage{multirow}
\usepackage{multicol}
\usepackage{makecell} 
\usepackage{booktabs}
\usepackage{epstopdf} 
\usepackage[locale=UK,detect-all=true]{siunitx} 

\usepackage[
ruled,
vlined]{algorithm2e}
\SetKwInput{KwData}{Input}
\SetKwInput{KwResult}{Output}
\SetKwComment{Comment}{$\triangleright$\ }{}
\SetNlSty{textbf}{}{:}

\title{WellPINN: Accurate Well Representation for Transient Fluid Pressure Diffusion in Subsurface Reservoirs with Physics-Informed Neural Networks}

\usepackage{authblk}

\author[1,2]{Linus Walter\thanks{Corresponding author: linus.walter@csic.es}}
\author[3]{Qingkai Kong}
\author[1]{Sara Hanson-Hedgecock}
\author[1]{Víctor Vilarrasa\thanks{Co-Corresponding author: victor.vilarrasa@csic.es}}

\affil[1]{Global Change Research Group (GCRG), IMEDEA, CSIC-UIB, Spain}
\affil[2]{Department of Civil and Environmental Engineering (DECA), Universitat Politècnica de Catalunya·BarcelonaTech (UPC), Barcelona, Spain}
\affil[3]{Lawrence Livermore National Laboratory, Livermore, USA}

\date{\today}

\begin{document}

\maketitle

\textbf{Keypoints:}
\begin{itemize}
    \item WellPINN is a sequential training workflow using physics-informed neural networks (PINNs) that solve transient fluid flow problems
    \item We accurately represent a well at a reservoir-scale domain spanning three order of magnitude in space dimension
    \item Domain decomposition and logarithmic scaling of time is required to accurately represent wells with PINNs that use tanh activation functions
\end{itemize}

\section*{Abstract}

Accurate representation of wells is essential for reliable reservoir characterization and simulation of operational scenarios in subsurface flow models. Physics-informed neural networks (PINNs) have recently emerged as a promising method for reservoir modeling, offering seamless integration of monitoring data and governing physical equations. However, existing PINN-based studies face major challenges in capturing fluid pressure near wells, particularly during the early stage after injection begins. To address this, we propose WellPINN, a modeling workflow that combines the outputs of multiple sequentially trained PINN models to accurately represent wells. This workflow iteratively approximates the radius of the equivalent well to match the actual well dimensions by decomposing the domain into stepwise shrinking subdomains with a simultaneously reducing equivalent well radius. Our results demonstrate that sequential training of superimposing networks around the pumping well is the first workflow that focuses on accurate inference of fluid pressure from pumping rates throughout the entire injection period, significantly advancing the potential of PINNs for inverse modeling and operational scenario simulations.
All data and code for this paper will be made openly available at \url{https://github.com/linuswalter/WellPINN}.

\pagebreak
\section*{Plain Language Summary}
Accurately representing wells is crucial when building reservoir models to simulate fluid flow. During well testing, it is particularly important to match applied flow rates and observed pressures to calibrate the model. A promising tool are physics-informed neural networks (PINNs) that flexibly combine measured data with physical laws.
However, existing PINNs often fail to capture the fluid pressure field near wells, especially immediately after injection begins.
To overcome this limitation, we developed a new workflow called WellPINN. It uses several PINN models in sequence. The first model is trained for the whole modeling domain, while each subsequent model focuses on a smaller area around the well. Simultaneously, we also reduce the dimensions of the representative well.
We demonstrate WellPINN on a test case involving a single pumping well in a two-dimensional reservoir. Our results show that using three PINNs in sequence can successfully model pressure changes from the reservoir boundary to the well center across multiple spatial scales.
WellPINN is the first PINN-based method to accurately predict pressure throughout the entire injection period. This opens new possibilities for applying PINNs in inverse modeling and planning real-world reservoir operations, including in more complex reservoir settings or with multiple wells.

\section{Introduction} \label{sec-intro}

Wells are fundamental for hydrogeological applications such as drinking water production, aquifer remediation, and harnessing of geothermal energy \parencite{Holting2013_Hydrogeologie}. 
During the exploration phase, well tests help characterize subsurface parameters \parencite{WatsonAT1994_ParameterSystemIdentification,Slotte2017_LectureNotesWelltesting,Vaezi2024_Bedretto_Stimulation_Model}, while later stages require simulations of fluid pressure responses to operational scenarios \parencite{Bentley2020_ReservoirModelingSimulation}. 
These simulations use forward models to estimate the spatiotemporal distribution of fluid pressure \(p(\mathbf{x},t)\) in response to injection and/or pumping flow rates \(q_\text{w}\). State-of-the-art for reservoir modeling are numerical solvers like \lq{}MOOSE Framework\rq{} \parencite{Giudicelli2024_MOOSE_3}, \lq{}OpenGeoSys\rq{} \parencite{Naumov2022_OpenGeoSys}, \lq{}CODE\_BRIGHT\rq{} \parencite{Code_Bright2023}, and \lq{}GEOS\rq{} \parencite{GeosSoftware2024}, which solve conservation equations of mass, energy, and momentum using finite element, difference, or volume methods \parencite{Bentley2020_ReservoirModelingSimulation}. These methods discretize the geological domain with meshes that are refined around wells to resolve steep pressure gradients. 
Such models are effective for simulations, but they often struggle to incorporate noisy field data and quantify uncertainties.
A potential alternative could be data-driven machine learning approaches that flexibly integrate data \parencite{SunJ2022_data_driven_ML_Groundwater}, but typically require more data than subsurface systems can provide \parencite{ShenC2023_Differentiable_ML}.

To overcome these limitations, hybrid models, such as physics-informed neural networks (PINNs) \parencite{Lagaris_1998_PINN,Raissi2019_PINN_Original} and theory-guided networks (TgNNs) \parencite{Karpatne2017_TheoryGuidedDataScience}, combine the strengths of physics-based and data-driven models by embedding residuals of the governing partial differential equations (PDEs) into the loss function via automatic differentiation \parencite{Baydin2015_AutomaticDifferentiationMachine}. 
PINNs have shown potential across disciplines \parencite{Karniadakis2021_Physics_Based_ML}, supported by recent advances in loss weighting \parencite{WangS2023_ExpertGuidePINN}, activation function optimization \parencite{Abbasi2023_PhysicalActivationFunctions,Huang2023_GaborPINN}, and hard constraints \parencite{Sukumar2021_PINN_Hard_Constraints,Taufik2024_Hard_constraints}. In geosciences, PINNs are increasingly used in reservoir modeling. Recent developments include multiphysics coupling \parencite{Haghighat2021_PINN_sequential_split}, two-phase flow \parencite{Hanna2022_PINNResidualbasedAdaptivityTwophase}, sharp material transitions \parencite{Lehmann2023_Mixed_p_v_formulation_for_PINN,Sarma2024_IPINNs_DomainDecomp}, and fractured domains \parencite{YanX2024_PINN_fractured_reservoir, AbbasiJ2025_PINN_in_fractured_core}. 
Despite these advances, model simplifications such as homogeneous materials \parencite{Haghighat2021_PINN_sequential_split}, stationary flow \parencite{Lehmann2023_Mixed_p_v_formulation_for_PINN}, or dense grids of observation points \parencite{Tartakovsky2020_PINN,WangN2020_PINN_subsurface_flow} remain common, reflecting the ongoing challenges in adapting PINNs for practical reservoir modeling. 

Apart from these idealizations, a major challenge for PINNs is accurate well representation in the reservoir model. Analytical solutions represent singularities like wells as point sources. This representation is not suitable for PINNs because Artificial Neural Networks (ANNs) perform best on smooth, continuous, similarly scaled variables \parencite{Rumelhart1986_MLP_Backrpopagation}. Therefore, most studies represent wells in PINNs with equivalent well functions to smoothen steep pressure gradients in the vicinity of the well \parencite{Cuomo2023_GroundwaterPINN, Soriano2021_AssessmentGroundwaterWell, Zhang2022_GWPINN, Teng2022_GreenFunction_ML, Huang2021_PINN_with_Source_Term, WangN2024_PICD_flow_inversion, LiuA2024_PINN_Wells_Reservoir}. 

A common equivalent well representation is the Gaussian function, whose integral always equals one and which is directly derivable \parencite{Teng2022_GreenFunction_ML,Huang2021_PINN_with_Source_Term}. 
In addition, \textcite{Cuomo2023_GroundwaterPINN} compared the model performance of a piecewise cubic function to a cosine function, with a better performance for the latter one. \textcite{LiuA2024_PINN_Wells_Reservoir} obtained good results for well representation with a Lorentz function, although it requires a numerical method for integral calculation. Other levels of complexity are added by the use of multiple injection wells \parencite{Cuomo2023_GroundwaterPINN,LiuA2024_PINN_Wells_Reservoir,Zhang2022_GWPINN} or a heterogeneous modeling domain \parencite{LiuA2024_PINN_Wells_Reservoir}. Incorporating hard constraints effectively improves model convergence by reducing the number of loss terms \parencite{Zhang2022_GWPINN} and also domain decomposition produces more accurate results \parencite{Teng2022_GreenFunction_ML}.
A major advancement is the implementation of \textcite{WangN2024_PICD_flow_inversion} that provided a complex inversion workflow for the estimation of the permeability field based on a physics-informed deep convolutional encoder-decoder network that infers the evolution of the pore pressure field \(p(\mathbf{x},t)\) by using multiple wells and a range of observation points, also at the location of the well.
The use of observation points for fluid pressure within the modeling domain is common among several studies \parencite{Zhang2022_GWPINN,WangN2024_PICD_flow_inversion}. However, we are concerned that the number of incorporated observations exceeds the usually available number of observations in the field, especially for deep subsurface projects due to the high drilling cost \parencite{Beckers2019_Drilling_Costs_geothermal}.
Although these studies demonstrate the general suitability of equivalent well functions, they often require a relatively large equivalent well radius, typically at least \SI{10}{\%} of the domain size, to achieve stable convergence \parencite{Teng2022_GreenFunction_ML, Zhang2022_GWPINN, LiuA2024_PINN_Wells_Reservoir,Cuomo2023_GroundwaterPINN}. Although this is sufficient to capture \(p(\mathbf{x},t)\) in the far field for late injection times, these approaches underestimate the well pressure \(p_\text{w}\) by up to 30\% during late injection times and completely miss early-time diffusion. This limitation reduces the reliability of the model for history matching based on \(q_\text{w}\).
These limitations motivate the need for a PINN workflow capable of capturing a realistic representation of the well with sparse data.

We introduce \lq{}WellPINN\rq{} to address the limited well representation capabilities of previous studies by proposing a sequential training workflow that combines multiple PINN models. We subdivide the modeling domain into nested subdomains, with each inner model refining the solution near the well using the equivalent radius of the previously trained model as its new boundary. We first train a PINN across the entire domain, providing a function \(\hat{p}(\mathbf{x},t)\) that captures \(p(\mathbf{x},t)\) in the far field, but remains inaccurate near the well. Subsequently, trained PINNs focus on smaller subdomains, progressively refining the pressure field near the well by using smaller equivalent well radii. We apply the WellPINN workflow to a two-dimensional modeling domain with a single injection well of radius \(r = \SI{10}{\cm}\), located at the center of a \SI{100}{m} by side square. Our results show an accurate inference of \(p_\text{w}\) throughout the entire injection period, including early injection times. By isolating the partial derivative terms for each PINN model, we ensure consistent training times and avoid accumulating computational overhead by iteratively increasing the number of PINN functions. Thus, our approach offers a scalable solution to the multiscale challenge of well representation in a large domain, posing an important step for future applications of PINNs in history matching of pumping well tests.

\section{WellPINN: A Sequential Modeling Workflow for Well Representation with PINN}\label{sec-methods}

\begin{figure}
    \centering
      \makebox[\textwidth][c]{%
    \includegraphics[width=1.2\linewidth]{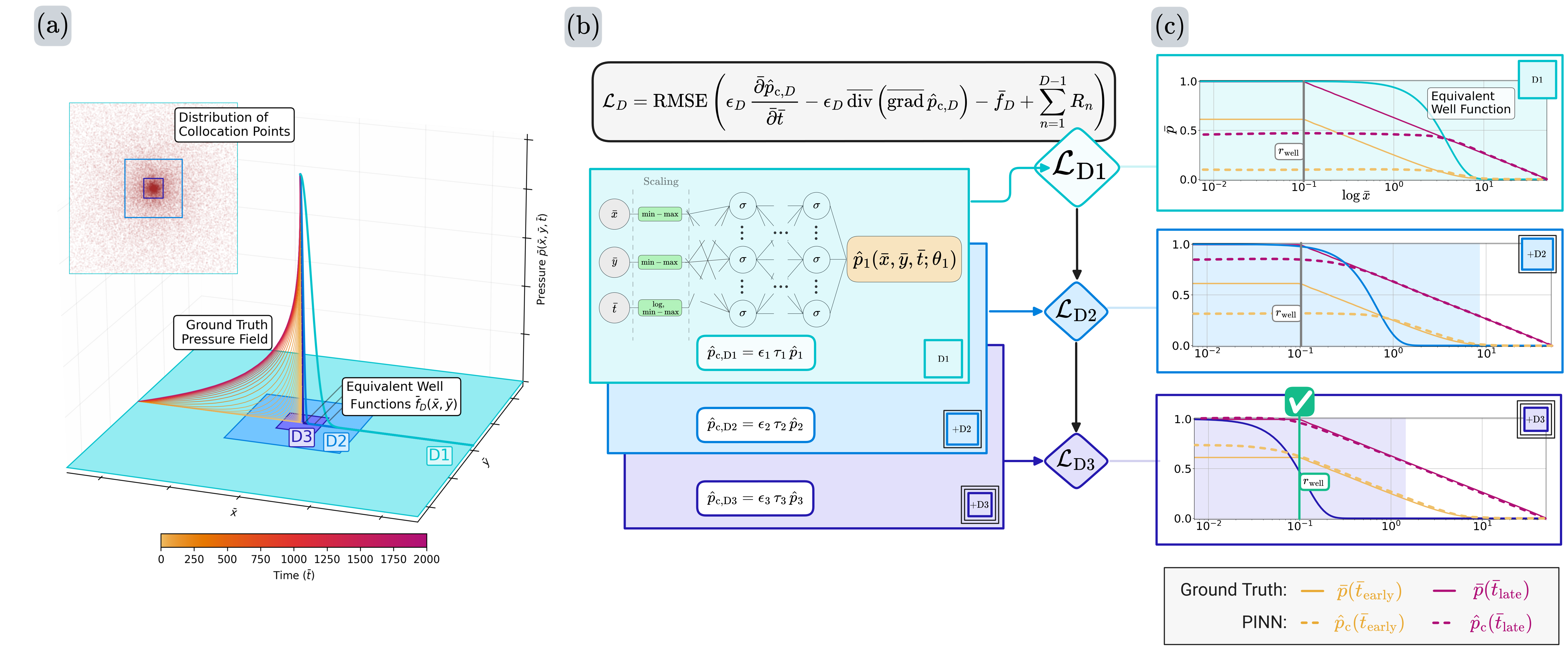}
    }
\caption{Sequential training workflow of our PINN model for transient fluid pressure diffusion from a single well in a 2D domain.
(a) Ground truth solution illustrating the diffusion of fluid pressure \(\bar{p}(\bar{x},\bar{y},\bar{t})\) from the injection well. The domain is decomposed into subdomains D1 (light blue), D2 (blue), and D3 (dark purple), with each assigned an equivalent well function \(\bar{f}_D(\bar{x},\bar{y})\). Collocation points are concentrated near the well (inset). 
(b) Schematic workflow of the training process showing the composite pressure solution \(\hat{p}_{\text{c},D}\) for each subdomain, constructed from the neural network output \(\hat{p}_D\), weighting factor \(\epsilon_D\), and a hard constraining approximate distance function \(\tau_D\). The training of \(\hat{p}_{\text{c}_D}\) is guided by the loss function \(\mathcal{L}_D\).
(c) In D1, the model \(\hat{p}_{c}\) infers the ground truth up to the equivalent well radius \(\bar{r}_\text{weq}\). In D2 and D3, a stepwise reduction of the training domain size \(\bar{r}_\text{max}\) (shaded area) and \(\bar{r}_\text{weq}\) (black line) improves each the inference of the ground truth \(\bar{p}(\bar{x},\bar{y},\bar{t})\) in the vicinity of the well.}
\label{figure-prj02-01-workflow}
\end{figure}

This study addresses the transient diffusion of fluid pressure \(p(x,y,t)\) in a 2D domain with a special focus on accurately representing a single well, as shown schematically in \cref{figure-prj02-01-workflow}a. The governing equation for \(p(x,y,t)\) reads

\begin{equation}
0=S_{\text{S}}\frac{\partial{p(x,y,t)}}{\partial{t}} - \mathrm{div}\, \left( \boldsymbol{K}(x,y) \,\mathrm{grad}\,p(x,y,t) \right) - \rho\,g f(x,y,t) ,
\label{eq-gov}
\end{equation}

where \(\boldsymbol{K}(x,y)\) is the hydraulic conductivity field, \(S_\text{s}\) is the specific storage coefficient, \(x\) and \(y\) are the Cartesian coordinates, \(t\) is time, and \(f(x,y,t)\) is a source/sink term per unit volume representing the well.  The gravity term is neglected in our case because gravity \(g\) is oriented perpendicular to our reservoir plane. We first introduce the relationship \(K(x,y) = k(x,y) \frac{\rho \times g}{\mu}\), where \(\rho \) is fluid density and \(\mu\) is fluid viscosity, and replace the specific storage \(S_{\text{S}}\) with its equivalent pressure formulation \(S_\text{S} = S'_\text{S}\,\rho \,g\). Then we cancel \(\rho\) and \(g\) in both terms and get

\begin{equation}
0=S'_\text{S}\frac{\partial{p}}{\partial{t}} - \frac{1}{\mu} \mathrm{div}\, \left( k(x,y)\, \mathrm{grad}\, p  \right) - f(x,y,t).
\label{eq-dim-trans}
\end{equation}
We determine \(S'_\text{S}\) from the definition for uniaxial specific storage as \(S'_\text{S} = \phi \kappa^{\textsf{f}}_{\text{R}} + (\alpha_{\text{B}}-\phi)\kappa^{\textsf{s}}_{\text{R}}\) where \(\phi\) is porosity, \(\kappa^{\textsf{f}}_{\text{R}}\) is the bulk modulus of the fluid \citep{Wang2000}. The source term $f({x},{y},{t})$ writes as
\begin{align}
f(x,y,t) &=    \frac{Q}{\,d} \, \frac{1}{2\pi\,\sigma^{2}} \exp \left({-\frac{0.5}{\sigma^{2}}\left( (x-x_{0})^{2} + (y-y_{0})^{2} \right)} \right),
\label{eq-source-term}
\end{align}
where \(d\) is the thickness of the reservoir, assumed as unity in the 2D model. The assigned values for the material parameters in \cref{eq-dim-trans} and \cref{eq-source-term}  are summarized in \cref{tabl_model_params}.
We introduce the following dimensionless parameters for the independent variables of space \(x,y\) and time \(t\) as well as for the dependent variables pressure \(p(x,y,t)\), permeability \(k(x,y)\) and the source term \(f(x,y,t)\)

\begin{align*}
&
\bar{x}=\frac{x}{{x}_\text{c}}\quad \bar{y}=\frac{y}{{y}_\text{c}}\quad \bar{t}=\frac{t}{{t}_\text{c}}\quad \\[1em]
&\bar{p}(\bar{x},\bar{y},\bar{t})=\frac{p(x,y,t)}{{p}_\text{c}} \quad \bar{k}(\bar{x},\bar{y})=\frac{k(x,y)}{k_{\text{c}}} \quad 
\bar{f}(\bar{x},\bar{y},\bar{t})=\frac{f(x,y,t)}{f_{\text{c}}}.
\end{align*}

We assume that \(y_\text{c}=x_\text{c}\) and replace each dimensional variable \(n\) with the respective dimensionless \(\bar{n}\), denoted by a bar, and characteristic counterpart \(n_{\text{c}}\), denoted by the subscript \textsubscript{c},  in \cref{eq-dim-trans}, resulting in

\begin{align*}
0&=  S'_{\text{s}} \, \frac{{p_c}}{t_c}\,\frac{\bar{\partial} \bar{p}}{\bar{\partial}\bar{t}}  - \frac{k_{\text{c}}}{\mu} \frac{p_{c}}{ x^{2}_{\text{c}} }  \overline{\mathrm{div}}\,\left( \bar{k} \;\overline{\mathrm{grad}}\,\bar{p}\right) - f_{\text{c}}\,\bar{f}(\bar{x},\bar{y},\bar{t}) \\[3ex]
{0}&{=  \overbrace{\underbrace{S'_{\text{s}} \, \frac{x^{2}_{\text{c}}\,\mu}{k_{\text{c}}\,t_{\text{c}}}}_{=1}}^{A}\,\frac{\bar{\partial} \bar{p}}{\bar{\partial}\bar{t}}  -   \overline{\mathrm{div}}\,\left( \bar{k} \;\overline{\mathrm{grad}}\,\bar{p}\right) - \overbrace{\underbrace{\frac{x^{2}_{\text{c}}\,\mu\,f_{\text{c}}}{k_{\text{c}} p_{\text{c}}}}_\mathrm{=1}}^{B} \bar{f}(\bar{x},\bar{y},\bar{t})}.
\end{align*}

The factors in the first term, \(A\), and the third term, \(B\), are assumed to be equal to one to maintain the order unity (\cite{Langtangen2016_Scaling_PDEs}).

We choose the parameters \(f_{c}\), \(p_{\text{c}}\) and \(k_{\text{c}}\) by their maximum parameters

\begin{align*}
x_{\text{c}}&= \SI{1}{m}  \\[1ex]
k_{\text{c}}&= k_\text{max}  \\[1ex]
f_{\text{c}}&=  Q_{\text{max}}   {\frac{1}{\SI{1}{m}}} \frac{1}{ 1 \,\mathrm{m^2}} e^{0} \qquad\left[ \mathrm{s^{-1}} \right],&&\\
\end{align*}
where we assume \(\sigma^2=1/2\pi\) (see \cref{eq-source-term}).

Therefore, we choose the remaining variables as the following:

\begin{align*}
p_{\text{c}} &=  \frac{x^{2}_{\text{c}} f_{\text{c}} \mu}{k_\text{c}} &t_{c} &= \frac{x^{2}_{\text{c}}\,S_\text{s}\,\mu}{\,k_{\text{c}}} 
\end{align*}

The dimensionless form of \cref{eq-gov}  reads therefore as

\begin{equation}
R = \frac{\bar{\partial} \hat{p}( \bar{x},\bar{y},\bar{t};\theta)}{\bar{\partial}\bar{t}}  -  \overline{\mathrm{div}}\left(\overline{\mathrm{grad}}\,\hat{p}( \bar{x},\bar{y},\bar{t};\theta)\right) - \bar{f}(\bar{x},\bar{y},\bar{t}),
\label{eq-gov-dimless}
\end{equation}
In addition to the dimensionless formulation, we represent the source/sink term similarly to \textcite{Zhang2022_GWPINN,Teng2022_GreenFunction_ML} with a bi-variate formulation of the Gaussian function as

\begin{align}
\bar{f}(\bar{x},\bar{y}) &=  \frac{1}{2 \pi \,\sigma^{2}} \times  \exp\left(-\frac{(\bar{x} - \mu_{x})^{2}+(\bar{y}-\mu_{y})^{2}}{2\sigma^{2}}\right),
\label{eq-point-source}
\end{align}

where \(\mu_{x}\) and \(\mu_{y}\) represent the location of the well within the domain.

We infer the dimensionless ground truth \(\bar{p}(\bar{x},\bar{y},\bar{t})\) using an ANN \(\hat{p}(\bar{x},\bar{y},\bar{t};\theta)\), where \(\hat{\cdot}\) denotes the ANN output, based on the multilayer perceptron \parencite{MinskyM1972_MLP,Rumelhart1986_MLP_Backrpopagation}, following the PINN framework \parencite{Lagaris_1998_PINN,Raissi2019_PINN_Original}. As shown in \cref{figure-prj02-01-workflow}b, the PINN receives three dimensionless input variables \(\bar{x}\), \(\bar{y}\), and \(\bar{t}\). Before being passed to the network, the spatial dimensions \(\bar{x}\) and \(\bar{y}\) are scaled with a min-max scaler, while time \(\bar{t}\) is log-scaled to account for the early-time nonlinearity in pressure diffusion. Due to smooth activation functions, \(\hat{p}( \bar{x},\bar{y},\bar{t};\theta)\) is continuous and therefore differentiable at all collocation points \(\left\{x^{i}_{\text{cp}},y^{i}_{\text{cp}},t^{i}_{\text{cp}}, p^{i}_{\text{cp}}  \right\}^{N_{\text{cp}}}_{i=1}\) by using automatic differentiation \parencite{Baydin2015_AutomaticDifferentiationMachine}. Next, we compute the residual \(R\) of \cref{eq-gov-dimless} and define the PINN loss as

\begin{align}
\mathcal{L} &= \mathcal{L}_\text{pde}= \text{RMSE} (R),
\label{eq-global-loss}
\end{align}

where the PDE loss \(\mathcal{L}_{\text{pde}}\) is the root mean squared error (RMSE) of \(R\) \parencite{Raissi2019_PINN_Original}. The optimizer adjusts the weights and biases of the network to minimize \(\mathcal{L}\) during training. 

As \cref{eq-global-loss} indicates, we use only one loss term in our workflow. That is because we do not use any observations of \(\bar{p}(\bar{x},\bar{y},\bar{t})\) for the model training, which would otherwise facilitate the training process. Secondly, we incorporate all additional constraints, such as boundary conditions and initial conditions, directly into a composite pressure function \(\hat{p}_\text{c}( \bar{x},\bar{y},\bar{t};\theta)\) that is written as

\begin{equation}
\hat{p}_\text{c}( \bar{x},\bar{y},\bar{t};\theta) = \hat{p}( \bar{x},\bar{y},\bar{t};\theta) \times \tau(\bar{x},\bar{y},\bar{t}),
\label{eq-composite-pressure}
\end{equation}

where \(\tau(\bar{x},\bar{y},\bar{t})\) is a piecewise approximate distance function. We adopt \(\tau(\bar{x},\bar{y},\bar{t})\) from \textcite{Sukumar2021_PINN_Hard_Constraints} and implement it as a hard constraint for our PINN model similar to \textcite{LuL2021_PINN_HardConstraints}, \textcite{RoyP2024_ExactEnforcementTemporal}, \textcite{LaiMC2023_HardConstraintPINNsInterface}, and \textcite{Zhang2022_GWPINN} as

\begin{equation}
    \tau(\bar{x},\bar{y},\bar{t}) = 
\begin{cases} {\Large
 \frac{(\bar{x}-\bar{x}_{\text{min}})(\bar{x}_\text{max}-\bar{x})(\bar{y}-\bar{y}_{\text{min}})(\bar{y}_\text{max}-\bar{y})\;\bar{t}}{0.5^{4}\times (\bar{x}_{\max}-\bar{x}_{\min})^{2} \times (\bar{y}_{\max}-\bar{y}_{\min})^{2}\;\bar{t}_{\text{max}}}}, & \text{if } |\bar{x}| \leq \bar{x}_{\max}\text{, }|\bar{y}| \leq \bar{y}_{\max} \\[1em]
0, &\text{if } |\bar{x}| > \bar{x}_{\max}\text{, }|\bar{y}| > \bar{y}_{\max}
\end{cases},
\label{eq-adf}
\end{equation}

where \(\bar{x}_{\max}\) and \(\bar{y}_{\max}\) are the domain boundaries and \(\bar{t}_{\max}\) is the maximum modeling time. \(\tau(\bar{x},\bar{y},\bar{t})\) ensures that \(\hat{p}_\text{c}( \bar{x},\bar{y},\bar{t};\theta)\) equals zero at any domain point beyond the domain boundary, as well as at the initial condition. In addition, \(\tau(\bar{x},\bar{y},\bar{t})\) is sufficiently partially differentiable for our diffusion problem, that is, twice differentiable over \(\bar{x}\) and once differentiable over \(\bar{t}\).

As introduced in \cref{sec-intro}, PINNs require a minimum ratio \(b > \SI{10}{\%}\), where \(b\) is the ratio between the equivalent well radius \(\bar{r}_\text{weq}\) and the domain extension \(\bar{x}_{\max}\). To not only meet this condition, but also ensure correct well representation, we introduce WellPINN, a sequential training approach for a series of PINNs (\cref{figure-prj02-01-workflow}). Here, we minimize \(\mathcal{L}\) stepwise by training each subsequent PINN model on a smaller subdomain, while using the output of the previously trained PINNs as the initial solution for the near field of the well, by adapting the approach of \textcite{WangY2024_MultistageNeuralNetworks}. 
For each training step \(D\), we calculate the equivalent well radius as

\begin{equation}
\bar{r}_\text{weq}=b^{D}\,\bar{x}_{\max,D},
\label{eq-r-well-eq}
\end{equation}

where we observe optimal regression results for \(b=0.17\) (see \cref{sec-discussion}). We set the subdomain size of each subsequent step to the equivalent radius of the previous step as \(\bar{x}_{\max,D} = \bar{r}_{\text{weq},D-1}\)  (see \cref{figure-prj02-01-workflow}c). The total number of subdomains \(N\) is defined as 
\begin{equation}
N > \frac{\log\left(\frac{\bar{r}_\text{w}}{\bar{x}_{\max}}\right)}{\log(b)}\qquad ,\quad N \in \mathbb{N},
\label{eq-nr-subdomains}
\end{equation}

where \(\bar{r}_\text{w}\) is the radius of the real well. We calculate the standard deviation for \cref{eq-point-source} as

\begin{equation}
\sigma_{D}=\frac{|\bar{r}_\text{weq} - \mu_{r}|} {\sqrt{-2 \,\ln(\beta)}},
\label{eq-std}
\end{equation}

where \(\beta = 10^{-2}\) is the target amplitude of the source function that defines \(\bar{r}_\text{weq}\).

Each pressure solution \(\hat{p}_{\text{c},D}\) is composed of all previously trained PINN outputs as

\begin{equation}
\hat{p}_\text{c,D}(\bar{x},\bar{y},\bar{t};\theta_{D}) =  \sum\limits^{D}_{n=1} \epsilon_{n} \times \tau_{n}(\bar{x},\bar{y},\bar{t}) \times \hat{p}_{n}( \bar{x},\bar{y},\bar{t};\theta_{n}),
\label{eq-p-comp}
\end{equation}

where \(\epsilon_{n}\) is a specific weighting factor \parencite{WangY2024_MultistageNeuralNetworks}. 
We assume \(\epsilon_D = 1\) at the initialization of the model and update the domain boundaries in \(\tau_{D}(\bar{x},\bar{y},\bar{t})\) for each step. As illustrated in \cref{figure-prj02-01-workflow}b and \cref{figure-prj02-01-workflow}c, training begins with the outer domain, where \(\hat{p}_{\text{c},1}\) approximates the far field. To correct deviations near \(\bar{r}_\text{weq}\), the residuals of previous steps are minimized in the following training loss, similar to the multistage PINN training of \textcite{WangY2024_MultistageNeuralNetworks}. 

Using the complete composite solution \(\hat{p}_{\text{c}}\) in the PDE residual would require evaluating all PINNs at each epoch. Since only the current network is trainable and the prior ones remain fixed, we reduce the computational cost by isolating the terms of each \(\hat{p}_{\text{c},D}\) and incorporating residuals from earlier stages \(R_n\) as precomputed terms 

\begin{equation}
\mathcal{L}_{D} = \text{RMSE}\left(  \epsilon_{D}\,\frac{\bar{\partial} \hat{p}_{\text{c},D}}{\bar{\partial}\bar{t}}  -  \epsilon_{D}\,  \overline{\mathrm{div}}\left(\overline{\mathrm{grad}}\,\hat{p}_{\text{c},D}\right) - \bar{f} +  \sum\limits^{D-1}_{n=1} R_{n} \right),
\label{eq-lossseq}
\end{equation}
where \(\sum^{D-1}_{n=1} R_n\) is computed at the beginning of each training stage \(D\). For an optimal scaling of the serial modeling steps, we update the specific weighting factor \(\epsilon_{I}\) at the beginning of each subsequent training step as

\begin{equation}
    \epsilon_{I} \,=\text{Mean} \left| \frac{ \bar{f} -  \sum\limits^{I-1}_{n=1} R_{n}}{ \bar{\partial} \hat{p}_{I}/ \bar{\partial}\bar{t}  -    \overline{\mathrm{div}}\left( \overline{\mathrm{grad}}\, \hat{p}_{I}  \right)}  \right|,
    \label{eq-weighting-factor}
\end{equation}
where \(\hat{p}\) is the yet untrained network of the respective subdomain. A full derivation of \cref{eq-lossseq} and \cref{eq-weighting-factor} is provided in Supplementary Information S1. The Training is performed sequentially using the Adam optimizer followed by L-BFGS \parencite{Kingma2015_Adam,LiuDC1989_LBFGS}. We summarize our training workflow \cref{algo-seq-training}.

\begingroup
\begin{algorithm}[H]
\caption{Sequential PINN Training for Accurate Well Representation}
\label{algo-seq-training}


Calculate the minimum number of subdomains
\(
N > \log\left( \frac{\bar{r}_\text{w}}{\bar{x}_{\max}} \right)/ \log(b), \quad N \in \mathbb{N}
\)

\For{$D = 1, 2, \dots, N$}{
    \eIf{$D = 1$}{
        $\bar{x}_{\max,D} = \bar{x}_{\max}$\;
    }{
        $\bar{x}_{\max,D}= \bar{r}_{\text{weq}, D-1}$
    }
    
    Compute equivalent well radius    \(    \bar{r}_{\text{weq}} \) from \cref{eq-r-well-eq}

Compute standard deviation of the equivalent well  \(\sigma_D  \) from \cref{eq-std}

Generate collocation points within subdomain in the boundaries of \(\bar{x}_{\min,D}<\bar{x}<\bar{x}_{\max,D}\) and  \(\bar{y}_{\min,D}<\bar{y}<\bar{y}_{\max,D}\).
    
Construct hard constraint function $\tau_D(\bar{x},\bar{y}, \bar{t})$ from \cref{eq-adf}
    
Construct composite solution assuming $\epsilon_D = 1$ from \cref{eq-p-comp}

Update weighting factor for current training step $\epsilon_D$ according to \cref{eq-weighting-factor}.

Construct residual loss \(\mathcal{L}_D\) as in \cref{eq-lossseq}.

Train network $\hat{p}_{\text{c},D}(\bar{x},\bar{y},\bar{t};\theta_D)$ by updating $\theta_D$ using the Adam optimizer
    
Update weighting factor \(\epsilon_D\)
    
Fine-tune $\hat{p}_{\text{c},D}(\bar{x},\bar{y},\bar{t};\theta_D)$ by updating $\theta_D$ using the L-BFGS optimizer
}
\KwResult{
Composite solution of fluid pressure as in \cref{eq-p-comp}
}
\end{algorithm}
\endgroup

\pagebreak

\section{Case Study for a 2D PINN with a Single Well}\label{sec-results}

\paragraph{Data Generation with a Numerical Model:}

We construct a synthetic training dataset by solving \cref{eq-gov} with the OpenGeoSys numerical solver \parencite{Naumov2022_OpenGeoSys}. We define a two-dimensional domain with the extensions of \(\SI{-50}{m}<x<\SI{50}{m}\) and \(\SI{-50}{m}<y<\SI{50}{m}\). We inject at a rate of \(q_\text{w}=\SI{9.9e-7}{\m\cubed\per\s\per\m}\) in a well with a radius of \(\bar{r}_\text{w}=\SI{1e-1}{\m}\) in the center of the domain. For all boundaries and for the initial condition boundaries, we define a Dirichlet boundary condition with \(p=\SI{0}{\MPa}\). The material parameters are assumed for a granite-like material with a homogeneous permeability distribution. 
The PINN has an architecture of \(4 \times [40]\) (4 layers, each layer has 40 neurons) with \(\texttt{tanh}\) activation functions, and we train each PINN model for \num{24000} training epochs. We provide a table with all material parameters and with the detailed parameterization of our PINN model in \cref{tabl_model_params}.
We run our PINN model on an NVIDIA RTX A4500 GPU. The training times for our three models are \SIlist{696;1317;1801}{\s}, respectively.
\begin{table}[h]
	\centering
	\caption{Overview of all operational parameters, material parameters of the numerical model and parameters of the PINN model for the 2D reservoir model with a central injection well} 
	\vspace{1mm}
	{\small\tabulinesep=1.5mm
		\begin{tabu} 
			{
				>{\raggedleft\arraybackslash}p{4cm} 
				|>{\raggedleft\arraybackslash}p{5cm} 
				>{\raggedright\arraybackslash}p{5cm}	
			}
			\toprule
			
			& \textbf{Parameter} & \textbf{Values} \\
			\midrule
			\textbf{Operational Parameters} & Volumetric Flow Rate & \({q}_\text{w} = \SI{9.9e-7}{\m\cubed\per\s\per\m} \)\\
			& & \\
			\midrule
			
			\textbf{Material parameters of the Numerical Model} & \\
			Fluid& Dynamic viscosity & \(\mu =\SI{1.006e-3}{\Pa \s}\) \\
			& Density & \( \varrho^{\textsf{f}}_{\text{R}} = \SI{998.2}{\kg \per\m\cubed}\) \\
			& Compressibility & \(\kappa^{\textsf{f}}_{\text{R}} = \SI{5.0e-10}{\per\Pa}\) \\
			Solid & Density & \(\varrho^{\textsf{s}}_\text{R}=  \SI{2690}{\kg \per \m\cubed}\) \\
			& Porosity & \(\phi = \num{0.01}\) \\
			& Biot-Willis coefficient & \(\alpha_\text{b} = 0.8\) \\
			& Compressibility & \(\kappa^{\textsf{s}}_\text{R} = \SI{4.89e-11}{\per \Pa}\) \\
			& Specific storage as pressure formulation & \(S'_{\text{s}}=\SI{4.36e-11}{\per\Pa}\) \\ 
			& Intrinsic permeability & \(k = \SI{1e-16}{\m\squared}\) \\
			& & \\
			\midrule
			\textbf{Parameters of the PINN} & ANN architecture & \(4 \times [40]\) \\
			& Activation Function & \texttt{tanh} \\
			& Output Activation Function & \texttt{softplus} \\
			& Learning Rate & Exponential reduction from \num{1e-2} to \num{5e-4} \\
			& Loss terms & \(\mathcal{L}_{\text{PINN}} = \mathcal{L}{\text{pde}} \)\\
			& Nr of Collocation Points & \num{50000} (see Supplementary Information S2) \\
			& Nr of Observation Points & 0 \\
			& Optimizer & Adam + L-BFGS \\
			& Training Epochs & \num{24000} per Training Sequence \\
			& Loss Term Evaluation Metric & RMSE \\
			& Nr of Subdomains & 3 \\
			& CP Refinement & Radial refinement following power distribution  \\
			& Time scaling & logarithmic with injection start at \(\bar{t}=\num{0.01}\) \\
			\bottomrule		\end{tabu}}
	\label{tabl_model_params}
\end{table}

\paragraph{Modeling Results:}

\begin{figure}[H]
\centering
\includegraphics[width=\linewidth]{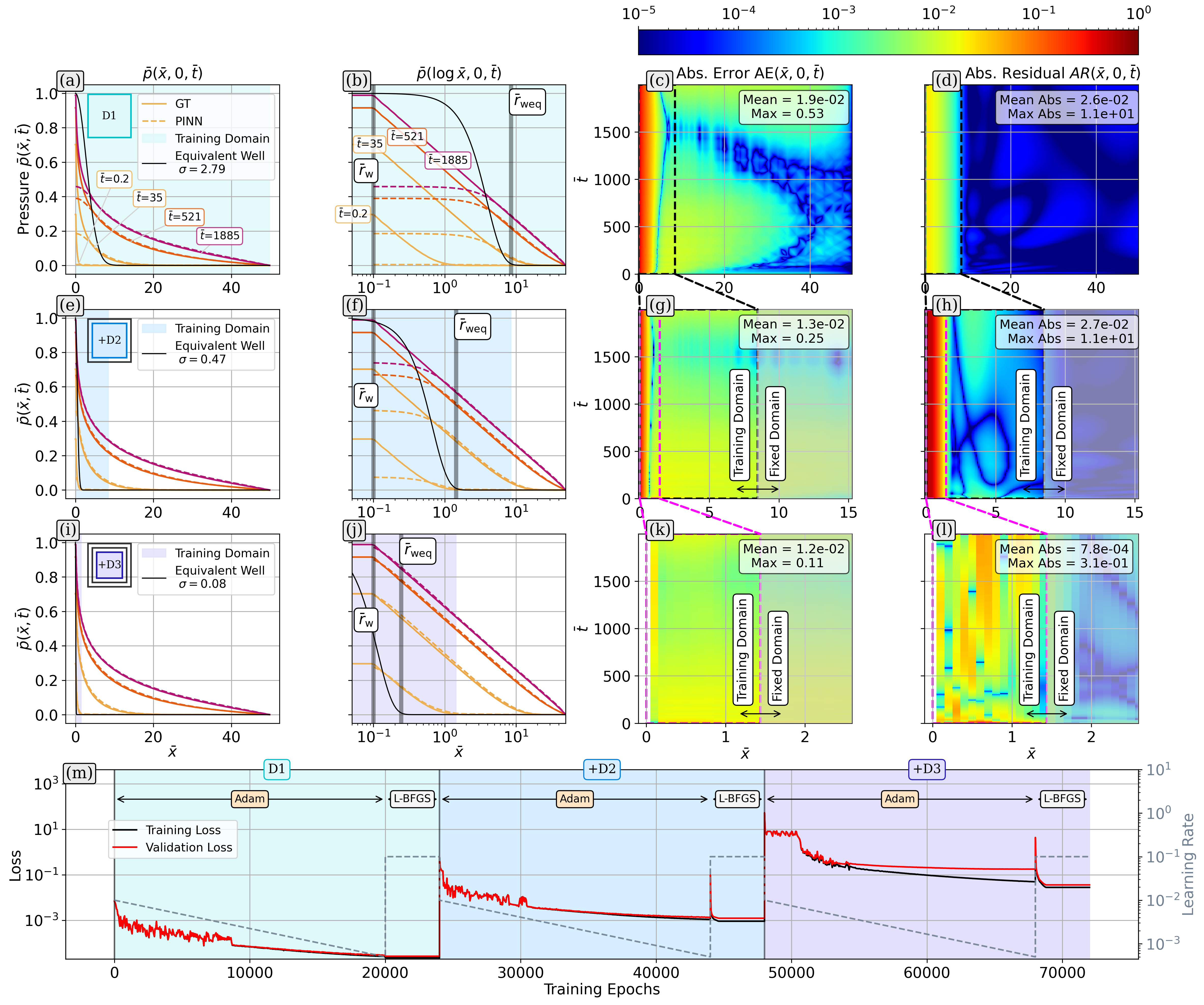}
\caption{Modeling results of PINN model for transient fluid pressure diffusion with a single well in a 2D domain at the cross section of \(\bar{y}=0\). (a,e,i) Comparison of the ground truth \(\bar{p}(\bar{x},0,\bar{t})\) versus the PINN output \(\hat{p}_\text{c,D}(\bar{x},0,\bar{t};\theta_{D})\) for timesteps \(\bar{t}=[0.2,35,521,1885]\) for a linear \(\bar{x}\) for the modeling steps D1, D2 and D3, respectively. (b,f,j) scale \(\bar{x}\) in log, where the real well radius \(\bar{r}_\text{w}\) and equivalent radius \(\bar{r}_\text{weq}\) are indicated. (c,g,k) Absolute error \(\text{AE}(\bar{x},0,\bar{t})\) for D1,D2, and D3 with a zoom towards the training domains for D2 and D3. (d,h,l) Absolute Residual \(\text{AR}(\bar{x},0,\bar{t})\) for D1,D2, and D3 with a zoom towards the training domains for D2 and D3. (m) Evolution of training loss and validation loss for D1, D2, and D3.}\label{figure-prj02-result_01}
\end{figure}

To demonstrate that our model fulfills the requirements outlined in \cref{sec-intro} for exact well representation, we present a detailed training visualizations in \cref{figure-prj02-result_01} with the first three rows showing the results from the three sequential PINN models. The third and fourth columns show the absolute error \(\text{AE}(\bar{x},0,\bar{t})\) and the residual \(\text{AR}(\bar{x},0,\bar{t})\), respectively. The fourth row shows the training and validation loss for the three domains. 

In the initial training stage D1 (\cref{figure-prj02-result_01}, row one), the cross section in \cref{figure-prj02-result_01}a indicates a strong agreement across most of the domain. However, \(\hat{p}_\text{c,D1}\) deviates from \(\bar{p}\) within the equivalent well. In particular, early timesteps (\(\bar{t}=0.2\) and \(\bar{t}=35\)) are not captured, as their diffusion fronts lie within \(\bar{r}_\text{weq}\). As shown in \cref{figure-prj02-result_01}c, the AE remains below \num{0.02} in most areas, but rises to \num{0.53} for \(\bar{x}<9\). The absolute residual (\(\text{AR}\)) in \cref{figure-prj02-result_01}d mirrors this trend, as it remains below \num{1e-4} in the far field, but increases abruptly within \(\bar{r}_\text{w}\), indicating difficulty in capturing the steep gradient of the source term at the well boundary. During D1, we therefore encounter two effects: a mismatch in the vicinity of the well and a mismatch for early timesteps. Both effects motivate our study to pursue the sequential training approach with a stepwise reduction of \(\bar{r}_\text{weq}\).

In training stage D2, the subdomain shrinks to the extent of the equivalent well of stage D1. Here, the cross sections (\cref{figure-prj02-result_01}e and \cref{figure-prj02-result_01}f) show an improved fit near \(\bar{r}_\text{w}\), and \cref{figure-prj02-result_01}g reveals that \(\text{AE}_{\max}\) is reduced by half to \num{0.25}. However, residual discrepancies remain within \(\bar{r}_\text{weq}\), especially at the source boundary ( \cref{figure-prj02-result_01}h). D3 further reduces \(\bar{r}_\text{weq}\) towards \(\bar{r}_\text{w}\), nearly eliminating the mismatch, as shown in  \cref{figure-prj02-result_01}i and  \cref{figure-prj02-result_01}j. Here, even early timesteps are captured, with the mean absolute error (MAE) reduced to \num{1.02e-2} and \(\text{AE}_{{\max}}\) being reduced from the initial \num{0.53} in D1 to \num{0.11} in D3. Similarly, the maximum value of \(AR\) reduces from \num{1.1e1} in D1 to \num{3.1e-1} in D3.

The loss evolution provides further information about the model training (\cref{figure-prj02-result_01}m). Each stage comprises \num{24000} epochs (Adam: \num{20000}, L-BFGS: \num{4000}). Training loss decreases in D1 from \num{1e-1} to \num{1e-4}. For the second training stage, we focus the training on the subdomain with the highest AE and highest \(\text{AR}\) values, resulting in an initially higher training loss in D2 with \num{7e-1}, which reduces to \num{1e-3}.
Similarly, training step D3 is initialized with large loss values, namely \num{3e1}, which is reduced by two orders of magnitude to \num{9e-2}. 
A divergence between training and validation loss appears in D3 during Adam optimization but resolves during L-BFGS. 
In total, the results for the application example in \cref{figure-prj02-result_01} show a satisfactory regression result, which implies that our sequential training is able to reduce the significant pressure mismatch of \(\text{AE}_{\max}=\num{0.53}\) to \(AE_{{\max}}=\num{0.11}\) throughout the injection time during two additional training iterations.

\begin{figure}[H]
    \centering
    \includegraphics[width=0.99\linewidth]{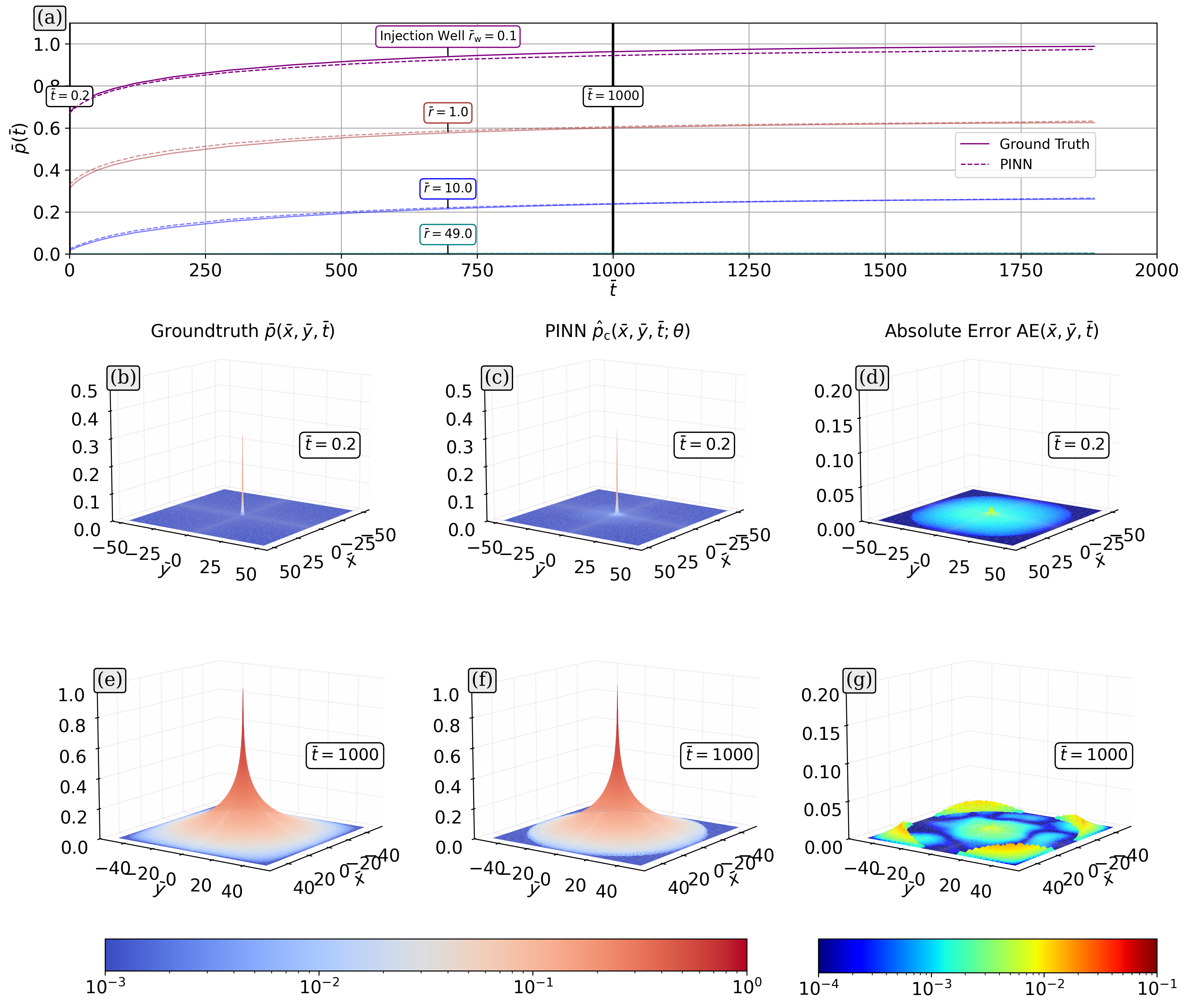}
    \caption{Pressure evolution at the well over the whole modeling domain. (a) Comparison of ground truth \(\bar{p}(\bar{x},\bar{y},\bar{t})\) and PINN output \(\hat{p}_\text{c}(\bar{x},\bar{y},\bar{t};\theta)\) over the injection time for the radii \(\bar{r}=[0.1,1.0,10,49]\). (b,c,d) 3D surfaces of \(\bar{p}(\bar{x},\bar{y},\bar{t})\), \(\hat{p}_\text{c}(\bar{x},\bar{y},\bar{t};\theta)\) and \(\text{AE}(\bar{x},\bar{y},\bar{t})\) for early time at \(\bar{t}=\num{0.2}\), respectively. (e,f,g) 3D surfaces of \(\bar{p}(\bar{x},\bar{y},\bar{t})\), \(\hat{p}_\text{c}(\bar{x},\bar{y},\bar{t};\theta)\) and \(\text{AE}(\bar{x},\bar{y},\bar{t})\) for late time at \(\bar{t}=\num{1000}\), respectively.}
    \label{figure-prj02-result_02}
\end{figure}

To further assess the inference of the well pressure, we plot the evolution of the pressure at different distances from the center of the domain \(\bar{r}\) (\cref{figure-prj02-result_02}a). The PINN model matches the ground truth sufficiently at \(\bar{r}_\text{w}\), with only minor deviations throughout injection time. Similar precision is seen at \(\bar{r}=1.0\) and \(10.0\), although slight mismatches occur until \(\bar{t}=500\). At the boundary \(\bar{r}=49\), the model infers the boundary condition nearly perfectly due to the hard constraint function \(\tau\). We also visualize 2D surface maps for early time \(\bar{t}=0.2\) (\cref{figure-prj02-result_02}b–d) and near-stationary state at \(\bar{t}=1000\) (\cref{figure-prj02-result_02}e–g). At \(\bar{t}=0.2\), the PINN model accurately captures the pressure front, while it slightly overestimates \(\bar{p}(\bar{x},\bar{y},\bar{t})\) outside of the front, as illustrated by the AE map in \cref{figure-prj02-result_01}d with a range of \(\num{0.001}<AE<\num{0.02}\).
At later times, discrepancies are most prominent at the domain corners, where \(\bar{p}(\bar{x},\bar{y},\bar{t})\) spreads non-radially while \(\hat{p}_\text{c}(\bar{x},\bar{y},\bar{t};\theta)\) remains rather radially shaped, leading to \(\text{AE}_{\max}=0.015\). In general, our PINN workflow achieves high accuracy, resolving the well with a ratio of \(b= \SI{0.17}{\%}\) per subdomain.

\section{Discussion}\label{sec-discussion}

\begin{figure}[H]
    \centering
    \includegraphics[width=0.99\linewidth]{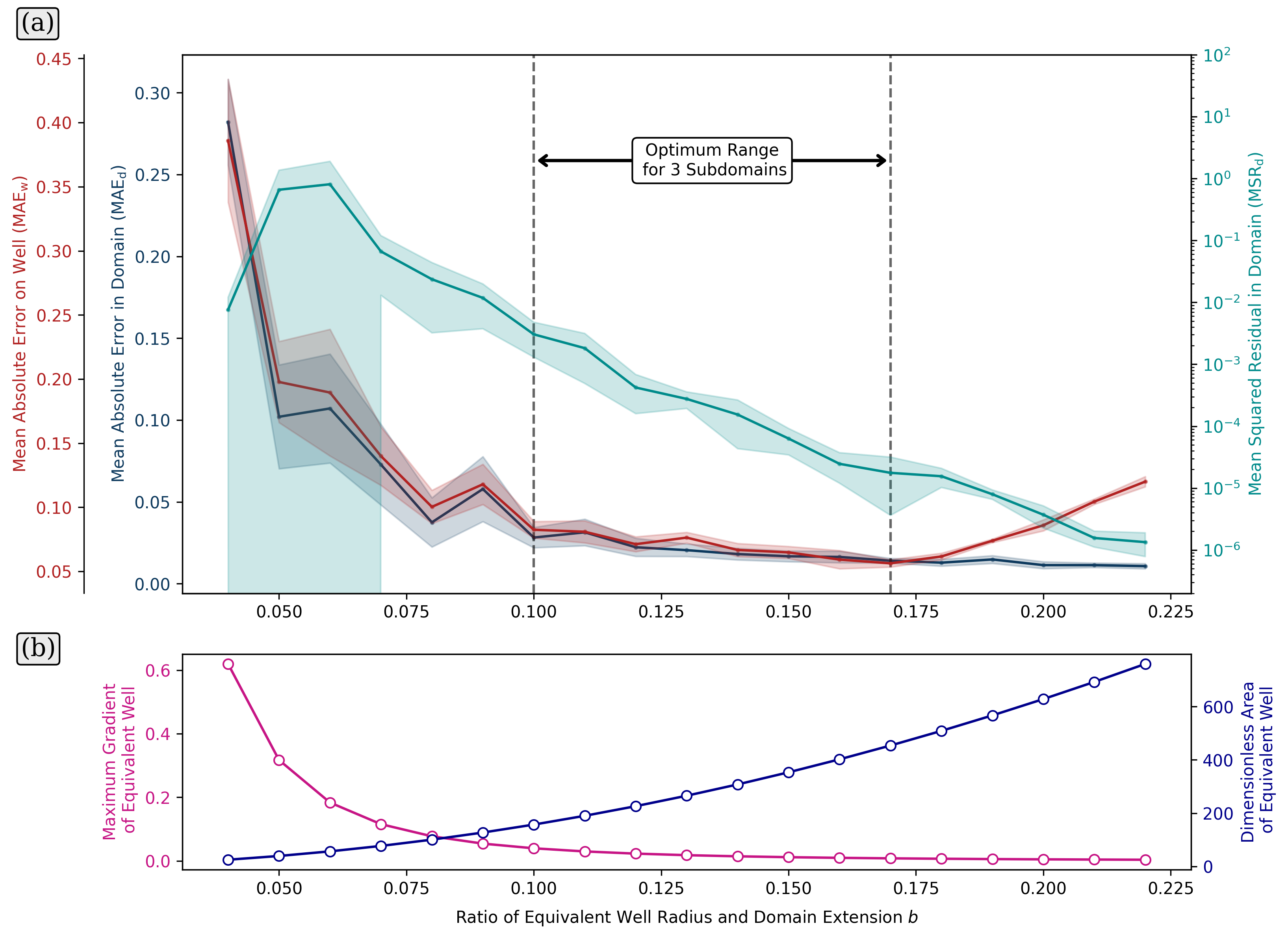}
    \caption{Parameter study for the well representation performance of the PINN model for the equivalent well ratio \(b\). (a) The mean absolute error is given for the well surface as \(\text{MAE}_{\text{w}}\) (dark red) and for the modeling domain as \(\text{MAE}_\text{d}\) (dark blue) and the mean squared residual \(\text{MSR}_\text{d}\) is also given for the modeling domain. We plot the mean (solid line) and the confidence interval (shaded area) for each metric. The optimum parameter range for having three subdomains is \(0.1<b<0.17\). (b) Evolution of the maximum gradient and the dimensionless area of the equivalent well function in dependency of \(b\), respectively.}
    \label{figure-prj02-result_03}
\end{figure}

The results demonstrate the optimal representation of the well in \(b=0.17 = \bar{r}_\text{weq}/\bar{x}_{\max}\) (\cref{sec-results}). 
To quantify the choice of \(b\), we conduct a parameter study for \(0.04 < b < 0.22\) (\cref{figure-prj02-result_03}), with five realizations per value of \(b\). We analyze the mean and the confidence interval for the mean absolute error on the well (\(\text{MAE}_\text{w}\)), as well as the mean absolute error (\(\text{MAE}_\text{d}\)), and the mean squared residual in the modeling domain (\(\text{MSR}_\text{d}\)).
All three metrics show optimal accuracy for \(0.1 < b < 0.17\). For \(b<0.1\), the increase of the maximum gradients in the source term in combination with the shrinking area of the equivalent well led to a worsening performance for all metrics (compare \cref{figure-prj02-result_03}a, \cref{figure-prj02-result_03}b and Supplementary Information S3). Adding more collocation points on the equivalent well has just a limited effect on reducing these errors, because the activation functions are not optimized for such a multi-scale problem. Interestingly, \(\text{MSR}_\text{d}\) decreases at \(b=0.04\), probably due to the optimizer ignoring a negligible source. For \(b>0.06\), the residual \(\text{MSR}_\text{d}\) constantly decreases, confirming that a PINN model with wider \(\bar{r}_\text{weq}\) converges more easily. For \(b>0.17\), \(\text{MAE}_\text{w}\) increases as \(\bar{r}_\text{weq}\) no longer overlaps well with \(\bar{r}_\text{w}\), also indicating that a fourth subdomain could solve this problem.

Our study finds that using a PINN model to represent a centimeter-scale pumping well in a reservoir-sized domain requires domain decomposition. 
Here, we propose a workflow that hard constraints the PINN output at the subdomain boundaries in combination with a sequential training towards the well. This setup improves model training by using only one loss term per PINN, while most importantly, it ensures strict continuity of the composite PINN function, unlike classical approaches which are more suitable for steep jumps at material boundaries (\cite{Shukla2021_PINN_Domain_decomp}). 
In our study, we avoided the use of observational data for \(\bar{p}(\bar{x},\bar{y},\bar{t})\) to demonstrate that our model infers the fluid pressure from physics alone.
However, a data loss term can be easily added, although an appropriate loss term weighting would be beneficial to obtain optimal results (\cite{WangS2023_ExpertGuidePINN, WangY2024_MultistageNeuralNetworks}). Furthermore, we find that logarithmic scaling of input time enhances early-time regression accuracy, a technique not previously applied to well representation in PINNs.


Compared to previous studies, our investigated value \(b=0.17\) for the ratio of the equivalent well radius to the domain extension aligns with the experience in the literature showing \(0.1<b<0.2\). Several previous studies enhanced model complexity with multiple wells (\cite{LiuA2024_PINN_Wells_Reservoir,Zhang2022_GWPINN}), heterogeneous domains (\cite{Zhang2022_GWPINN}), or compared model performance for various equivalent well functions (\cite{Cuomo2023_GroundwaterPINN}), achieving reasonable inference of \(\bar{p}(\bar{x},\bar{y},\bar{t})\) for the far field and for late time steps. We advance this current state-of-the-art with our WellPINN workflow (\cref{sec-methods}) that infers \(\bar{p}(\bar{x},\bar{y},\bar{t})\) in the vicinity of the well, proven in the exemplary case study in \cref{sec-results}, and our parameter study for feasible \(b\) values in \cref{sec-discussion}. Therefore, we provide an essential step for developing PINNs as versatile reservoir simulators for forward and inverse modeling. 
Although we test the application of WellPINN solely for a single injection well and constant permeability, we are confident that our workflow is extendable to a heterogeneous permeability field. For adding multiple wells, our current approach of a sequential training for each well might become computationally expensive. Instead of decomposing the domain for each well, future investigations could test sequential training with a stepwise reduced \(\bar{r}_\text{weq}\) and a frequency scaling factor for the activation functions in each \(\hat{p}_\text{c,D}(\bar{x},\bar{y},\bar{t};\theta_{D})\), as proposed by \citet{WangY2024_MultistageNeuralNetworks} for high-frequency domains. Another alternative would be to adapt the \lq{}hp-VPINN\rq{}-method of \citet{Kharazmi2021_hp_VPINN} for this purpose.
Although this multi-well modeling might require more investigation, we are confident that the inverse modeling, namely the determination of \(k_\text{mean}\) by matching \(p_\text{w}\) and \(q_\text{w}\) in the Darcy equation, is straightforwardly achievable with our WellPINN workflow.

\section{Conclusions}\label{sec-conclusions}
Accurately representing wells remains a key challenge in reservoir modeling, particularly for history matching of flow rate and well pressure during well tests. Although PINNs are becoming increasingly popular for these applications, existing studies have not yet explored how to reliably infer the entire fluid pressure field for a given flow rate. To close this gap, we propose the WellPINN workflow, a sequential PINN-based algorithm that decomposes the modeling domain and trains progressively from the far field toward the near field of the well. Each PINN is trained with a decreasing equivalent well radius \(\bar{r}_\text{weq}\), using the previous solution as initialization. Continuity of the composite PINN solution is achieved by hard constraining each PINN at the subdomain boundary. This approach results in a final equivalent well radius that matches the real well radius. In our case study, three PINNs are combined to model a well of \({r}_\text{w}=\SI{10}{\cm}\) in a domain of \(x_{\max}=\SI{50}{\m}\), while scaling time logarithmically. Our results demonstrate that WellPINN enables an accurate inference of \(p_\text{w}\), including at the early time after the start of injection. Our parameter study identifies \(b=0.17\) as the optimal ratio between the equivalent well radius \(\bar{r}_{\text{weq},D}\) and the subdomain extension \(\bar{x}_{\text{max},D}\). This result quantifies a trend noted in the literature: centimeter-scale wells cannot be resolved within large domains using a single PINN with standard tanh activation functions. Although further development is needed to optimize our WellPINN workflow for multiwell configurations, our study marks a relevant step toward establishing PINNs as a flexible reservoir simulator for the field of subsurface flow modeling.

%
%

\section{CRediT authorship contribution statement}

\textbf{Linus Walter:} Conceptualization, Formal analysis, Investigation, Methodology, Software, Visualization, Writing – original draft. \textbf{Qingkai Kong:} Methodology, Software, Supervision, Writing – review \& editing. \textbf{Sara Hanson-Hedgecock:} Supervision, Writing – review \& editing. \textbf{Víctor Vilarrasa:} Conceptualization, Funding acquisition, Methodology, Project administration, Resources, Supervision, Writing – review \& editing

\section*{Open Research Section} \label{sec:open-research}

We generate our data using the OpenGeoSys (OGS) version 6.4.3 open source code by \textcite{Naumov2022_OpenGeoSys} available at \url{https://www.opengeosys.org/}.
The PINN model is based on the pytorch version 2.3.1 of \textcite{pytorch_v231} openly available at \url{https://pytorch.org/}. Our sampling method for the collocation points is inspired by the SciANN Python package of \textcite{Haghighat2021_SciANNKerasTensorFlow}.
We created all figures with Matplotlib version 3.9.1 \parencite{HunterJD2007_Matplotlib,matplotlib_v291}, available under the Matplotlib license at \url{https://matplotlib.org/}.
All data sets are produced with numpy version 2.0.0 by \textcite{HarrisCR2020_Numpy} available at \url{https://pypi.org/project/numpy/} and with pandas version 2.2.2 by \textcite{pandas_v222} available at \url{https://pandas.pydata.org/}.

The input data includes the configuration file and the mesh for the numerical model as well as the output of the OGS model as a plain text file in CSV format.
All relevant input data and Python code necessary to reproduce our results are licensed under the GNU General Public License 3 or later (GPL-3.0). 
and will be made openly available at \url{https://github.com/linuswalter/WellPINN} following the journal publication of this manuscript.

\section*{Acknowledgments}

LW, SHH and VV acknowledge funding from the European Research Council (ERC) under the European Union’s Horizon 2020 Research and Innovation Program through the Starting Grant GEoREST (www.georest.eu) under Grant Agreement No. 801809. IMEDEA is an accredited "Maria de Maeztu Excellence Unit" (Grant CEX2021-001198, funded by MICIU/AEI/10.13039/501100011033). The work of QK is under the auspices of the US Department of Energy by Lawrence Livermore National Laboratory under contract DE-AC52 07NA27344. This paper is published under LLNL-JRNL-2006105.
We also express our gratitude to our collaborator Francesco Parisio for his contributions to the conceptualization of this project.




\printbibliography

@article{Abbasi2023_PhysicalActivationFunctions,
  title = {Physical {{Activation Functions}} ({{PAFs}}): {{An Approach}} for {{More Efficient Induction}} of {{Physics}} into {{Physics-Informed Neural Networks}} ({{PINNs}})},
  author = {Abbasi, Jassem and Andersen, P{\aa}l {\O}steb{\o}},
  year = {2023},
  pages = {26},
  doi = {10.48550/arXiv.2205.14630},
  langid = {english}
}

@article{AbbasiJ2025_PINN_in_fractured_core,
  title = {History-{{Matching}} of Imbibition Flow in Fractured Porous Media {{Using Physics-Informed Neural Networks}} ({{PINNs}})},
  author = {Abbasi, Jassem and Moseley, Ben and Kurotori, Takeshi and Jagtap, Ameya D. and Kovscek, Anthony R. and Hiorth, Aksel and {\O}steb{\o} Andersen, P{\aa}l},
  year = {2025},
  month = mar,
  journal = {Computer Methods in Applied Mechanics and Engineering},
  volume = {437},
  pages = {117784},
  issn = {00457825},
  doi = {10.1016/j.cma.2025.117784},
  url = {https://linkinghub.elsevier.com/retrieve/pii/S0045782525000568},
  urldate = {2025-02-05},
  langid = {english}
}

@article{Baydin2015_AutomaticDifferentiationMachine,
  title = {Automatic Differentiation in Machine Learning: A Survey},
  shorttitle = {Automatic Differentiation in Machine Learning},
  author = {Baydin, Atilim Gunes and Pearlmutter, Barak A. and Radul, Alexey Andreyevich and Siskind, Jeffrey Mark},
  year = {2015},
  publisher = {arXiv},
  doi = {10.48550/ARXIV.1502.05767},
  url = {https://arxiv.org/abs/1502.05767},
  urldate = {2024-01-17},
  copyright = {arXiv.org perpetual, non-exclusive license}
}

@article{Beckers2019_Drilling_Costs_geothermal,
  title = {{{GEOPHIRES}} v2.0: Updated Geothermal Techno-Economic Simulation Tool},
  shorttitle = {{{GEOPHIRES}} v2.0},
  author = {Beckers, Koenraad F. and McCabe, Kevin},
  year = {2019},
  month = dec,
  journal = {Geothermal Energy},
  volume = {7},
  number = {1},
  pages = {5},
  issn = {2195-9706},
  doi = {10.1186/s40517-019-0119-6},
  url = {https://geothermal-energy-journal.springeropen.com/articles/10.1186/s40517-019-0119-6},
  urldate = {2024-05-03},
  langid = {english}
}

@incollection{Bentley2020_ReservoirModelingSimulation,
  title = {Reservoir {{Modeling}} and {{Simulation}}},
  booktitle = {Encyclopedia of {{Petroleum Geoscience}}},
  author = {Bentley, Mark},
  editor = {Sorkhabi, Rasoul},
  year = {2020},
  pages = {1--13},
  publisher = {Springer International Publishing},
  address = {Cham},
  doi = {10.1007/978-3-319-02330-4_233-1},
  url = {http://link.springer.com/10.1007/978-3-319-02330-4_233-1},
  urldate = {2024-04-24},
  isbn = {978-3-319-02330-4},
  langid = {english}
}

@misc{Code_Bright2023,
  title = {{{CODE}}\_{{BRIGHT}}},
  author = {Olivella, Sebastia and Vaunat, Jean and {Rodriguez-Dono}, Alfonso},
  year = {2023},
  url = {https://deca.upc.edu/en/projects/code_bright},
  howpublished = {Universitat Polit{\`e}cnica de Catalunya}
}

@article{Cuomo2023_GroundwaterPINN,
  title = {Solving Groundwater Flow Equation Using Physics-Informed Neural Networks},
  author = {Cuomo, Salvatore and De Rosa, Mariapia and Giampaolo, Fabio and Izzo, Stefano and Schiano Di Cola, Vincenzo},
  year = {2023},
  month = sep,
  journal = {Computers \& Mathematics with Applications},
  volume = {145},
  pages = {106--123},
  issn = {08981221},
  doi = {10.1016/j.camwa.2023.05.036},
  url = {https://linkinghub.elsevier.com/retrieve/pii/S0898122123002602},
  urldate = {2024-07-09},
  langid = {english}
}

@misc{GeosSoftware2024,
  title = {{{GEOS}} Simulation Framework},
  author = {Castelletto, Nicola and Corbett, Benjamin and Cremon, Matthias and Fu, Pengcheng and Gross, Herv{\'e} and Hamon, Fran{\c c}ois and Huang, Jixiang and Klevtsov, Sergey and Lapene, Alexandre and Mazuyer, Antoine and Semnani, Shabnam and Settgast, Randolph and Sherman, Christopher and Vargas, Arturo and White, Joshua A. and White, Christopher},
  year = {2024},
  url = {https://github.com/libgeos/geos},
  howpublished = {Lawrence Livermore National Laboratory}
}

@article{Giudicelli2024_MOOSE_3,
  title = {3.0 - {{MOOSE}}: {{Enabling}} Massively Parallel Multiphysics Simulations},
  shorttitle = {3.0 - {{MOOSE}}},
  author = {Giudicelli, Guillaume and Lindsay, Alexander and Harbour, Logan and Icenhour, Casey and Li, Mengnan and Hansel, Joshua E. and German, Peter and Behne, Patrick and Marin, Oana and Stogner, Roy H. and Miller, Jason M. and Schwen, Daniel and Wang, Yaqi and Munday, Lynn and Schunert, Sebastian and Spencer, Benjamin W. and Yushu, Dewen and Recuero, Antonio and Prince, Zachary M. and Nezdyur, Max and Hu, Tianchen and Miao, Yinbin and Jung, Yeon Sang and Matthews, Christopher and Novak, April and Langley, Brandon and Truster, Timothy and Nobre, Nuno and Alger, Brian and Andrs, David and Kong, Fande and Carlsen, Robert and Slaughter, Andrew E. and Peterson, John W. and Gaston, Derek and Permann, Cody},
  year = {2024},
  month = may,
  journal = {SoftwareX},
  volume = {26},
  pages = {101690},
  issn = {23527110},
  doi = {10.1016/j.softx.2024.101690},
  url = {https://linkinghub.elsevier.com/retrieve/pii/S235271102400061X},
  urldate = {2025-03-27},
  langid = {english}
}

@article{Haghighat2021_PINN_sequential_split,
  title = {Physics-Informed Neural Network Simulation of Multiphase Poroelasticity Using Stress-Split Sequential Training},
  author = {Haghighat, Ehsan and Amini, Danial and Juanes, Ruben},
  year = {2021},
  month = oct,
  journal = {arXiv:2110.03049 [cs]},
  eprint = {2110.03049},
  primaryclass = {cs},
  doi = {10.1016/j.cma.2022.115141},
  url = {http://arxiv.org/abs/2110.03049},
  urldate = {2022-04-13},
  archiveprefix = {arXiv},
  langid = {english}
}

@article{Haghighat2021_SciANNKerasTensorFlow,
  title = {{{SciANN}}: {{A Keras}}/{{TensorFlow}} Wrapper for Scientific Computations and Physics-Informed Deep Learning Using Artificial Neural Networks [{{Software}}]},
  author = {Haghighat, Ehsan and Juanes, Ruben},
  year = {2021},
  journal = {Computer Methods in Applied Mechanics and Engineering},
  volume = {373},
  eprint = {2005.08803},
  issn = {00457825},
  doi = {10.1016/j.cma.2020.113552},
  archiveprefix = {arXiv}
}

@article{Hanna2022_PINNResidualbasedAdaptivityTwophase,
  title = {Residual-Based Adaptivity for Two-Phase Flow Simulation in Porous Media Using {{Physics-informed Neural Networks}}},
  author = {Hanna, John M. and Aguado, Jos{\'e} V. and {Comas-Cardona}, Sebastien and Askri, Ramzi and Borzacchiello, Domenico},
  year = {2022},
  month = jun,
  journal = {Computer Methods in Applied Mechanics and Engineering},
  volume = {396},
  pages = {115100},
  issn = {00457825},
  doi = {10.1016/j.cma.2022.115100},
  url = {https://linkinghub.elsevier.com/retrieve/pii/S004578252200295X},
  urldate = {2024-02-22},
  langid = {english}
}

@article{HarrisCR2020_Numpy,
  title = {Array Programming with {{NumPy}}},
  author = {Harris, Charles R. and Millman, K. Jarrod and Van Der Walt, St{\'e}fan J. and Gommers, Ralf and Virtanen, Pauli and Cournapeau, David and Wieser, Eric and Taylor, Julian and Berg, Sebastian and Smith, Nathaniel J. and Kern, Robert and Picus, Matti and Hoyer, Stephan and Van Kerkwijk, Marten H. and Brett, Matthew and Haldane, Allan and Del R{\'i}o, Jaime Fern{\'a}ndez and Wiebe, Mark and Peterson, Pearu and {G{\'e}rard-Marchant}, Pierre and Sheppard, Kevin and Reddy, Tyler and Weckesser, Warren and Abbasi, Hameer and Gohlke, Christoph and Oliphant, Travis E.},
  year = {2020},
  month = sep,
  journal = {Nature},
  volume = {585},
  number = {7825},
  pages = {357--362},
  publisher = {{Springer Science and Business Media LLC}},
  issn = {0028-0836, 1476-4687},
  doi = {10.1038/s41586-020-2649-2},
  url = {https://www.nature.com/articles/s41586-020-2649-2},
  urldate = {2025-04-30},
  copyright = {https://creativecommons.org/licenses/by/4.0},
  langid = {english}
}

@book{Holting2013_Hydrogeologie,
  title = {Hydrogeologie},
  author = {H{\"o}lting, Bernward and Coldewey, Wilhelm Georg},
  year = {2013},
  volume = {66},
  publisher = {Spektrum Akademischer Verlag},
  address = {Heidelberg},
  doi = {10.1007/978-3-8274-2354-2},
  url = {http://link.springer.com/10.1007/978-3-8274-2354-2},
  isbn = {978-3-8274-2353-5}
}

@misc{Huang2021_PINN_with_Source_Term,
  title = {Solving {{Partial Differential Equations}} with {{Point Source Based}} on {{Physics-Informed Neural Networks}}},
  author = {Huang, Xiang and Liu, Hongsheng and Shi, Beiji and Wang, Zidong and Yang, Kang and Li, Yang and Weng, Bingya and Wang, Min and Chu, Haotian and Zhou, Jing and Yu, Fan and Hua, Bei and Chen, Lei and Dong, Bin},
  year = {2021},
  number = {arXiv:2111.01394},
  eprint = {2111.01394},
  primaryclass = {physics},
  publisher = {arXiv},
  url = {http://arxiv.org/abs/2111.01394},
  urldate = {2023-02-20},
  archiveprefix = {arXiv},
  langid = {english}
}

@article{Huang2023_GaborPINN,
  title = {{{GaborPINN}}: {{Efficient Physics-Informed Neural Networks Using Multiplicative Filtered Networks}}},
  shorttitle = {{{GaborPINN}}},
  author = {Huang, Xinquan and Alkhalifah, Tariq},
  year = {2023},
  journal = {IEEE Geoscience and Remote Sensing Letters},
  volume = {20},
  pages = {1--5},
  issn = {1545-598X, 1558-0571},
  doi = {10.1109/LGRS.2023.3330774},
  url = {https://ieeexplore.ieee.org/document/10310254/},
  urldate = {2024-05-14},
  copyright = {https://ieeexplore.ieee.org/Xplorehelp/downloads/license-information/IEEE.html}
}

@article{HunterJD2007_Matplotlib,
  title = {Matplotlib: {{A 2D Graphics Environment}} [{{Software}}]},
  shorttitle = {Matplotlib},
  author = {Hunter, John D.},
  year = {2007},
  journal = {Computing in Science \& Engineering},
  volume = {9},
  number = {3},
  pages = {90--95},
  issn = {1521-9615},
  doi = {10.1109/MCSE.2007.55},
  url = {http://ieeexplore.ieee.org/document/4160265/},
  urldate = {2024-09-15},
  copyright = {https://ieeexplore.ieee.org/Xplorehelp/downloads/license-information/IEEE.html}
}

@article{Karniadakis2021_Physics_Based_ML,
  title = {Physics-Informed Machine Learning},
  author = {Karniadakis, George Em and Kevrekidis, Ioannis G and Lu, Lu and Perdikaris, Paris and Wang, Sifan and Yang, Liu},
  year = {2021},
  journal = {Nature Reviews Physics},
  volume = {3},
  number = {6},
  pages = {422--440},
  issn = {25225820},
  doi = {10.1038/s42254-021-00314-5},
  isbn = {0123456789}
}

@article{Karpatne2017_TheoryGuidedDataScience,
  title = {Theory-{{Guided Data Science}}: {{A New Paradigm}} for {{Scientific Discovery}} from {{Data}}},
  shorttitle = {Theory-{{Guided Data Science}}},
  author = {Karpatne, Anuj and Atluri, Gowtham and Faghmous, James H. and Steinbach, Michael and Banerjee, Arindam and Ganguly, Auroop and Shekhar, Shashi and Samatova, Nagiza and Kumar, Vipin},
  year = {2017},
  month = oct,
  journal = {IEEE Transactions on Knowledge and Data Engineering},
  volume = {29},
  number = {10},
  pages = {2318--2331},
  issn = {1041-4347},
  doi = {10.1109/TKDE.2017.2720168},
  url = {http://ieeexplore.ieee.org/document/7959606/},
  urldate = {2023-11-03}
}

@article{Kharazmi2021_hp_VPINN,
  title = {Hp-{{VPINNs}}: {{Variational}} Physics-Informed Neural Networks with Domain Decomposition},
  author = {Kharazmi, Ehsan and Zhang, Zhongqiang and Karniadakis, George E.M.},
  year = {2021},
  journal = {Computer Methods in Applied Mechanics and Engineering},
  volume = {374},
  eprint = {2003.05385},
  pages = {1--21},
  issn = {00457825},
  doi = {10.1016/j.cma.2020.113547},
  archiveprefix = {arXiv}
}

@article{Kingma2015_Adam,
  title = {Adam: {{A}} Method for Stochastic Optimization},
  author = {Kingma, Diederik P and Ba, Jimmy Lei},
  year = {2015},
  journal = {3rd International Conference on Learning Representations, ICLR 2015 - Conference Track Proceedings},
  eprint = {1412.6980},
  pages = {1--15},
  archiveprefix = {arXiv}
}

@article{Lagaris_1998_PINN,
  title = {Artificial Neural Networks for Solving Ordinary and Partial Differential Equations},
  author = {Lagaris, I.E. and Likas, A. and Fotiadis, D.I.},
  year = {1998},
  journal = {IEEE Transactions on Neural Networks},
  volume = {9},
  number = {5},
  pages = {987--1000},
  publisher = {{Institute of Electrical and Electronics Engineers (IEEE)}},
  issn = {1045-9227},
  doi = {10.1109/72.712178},
  url = {http://dx.doi.org/10.1109/72.712178}
}

@misc{LaiMC2023_HardConstraintPINNsInterface,
  title = {The {{Hard-Constraint PINNs}} for {{Interface Optimal Control Problems}}},
  author = {Lai, Ming-Chih and Song, Yongcun and Yuan, Xiaoming and Yue, Hangrui and Zeng, Tianyou},
  year = {2023},
  publisher = {arXiv},
  doi = {10.48550/ARXIV.2308.06709},
  url = {https://arxiv.org/abs/2308.06709},
  urldate = {2024-07-29},
  copyright = {arXiv.org perpetual, non-exclusive license}
}

@book{Langtangen2016_Scaling_PDEs,
  title = {Scaling of {{Differential Equations}}},
  author = {Langtangen, Hans Petter and Pedersen, Geir K.},
  year = {2016},
  publisher = {Springer International Publishing},
  address = {Cham},
  doi = {10.1007/978-3-319-32726-6},
  url = {http://hplgit.github.io/scaling-book/doc/pub/book/html/._scaling-book3000.html},
  urldate = {2022-08-05},
  isbn = {978-3-319-32725-9},
  langid = {english}
}

@article{Lehmann2023_Mixed_p_v_formulation_for_PINN,
  title = {A Mixed Pressure-Velocity Formulation to Model Flow in Heterogeneous Porous Media with Physics-Informed Neural Networks},
  author = {Lehmann, Fran{\c c}ois and Fahs, Marwan and Alhubail, Ali and Hoteit, Hussein},
  year = {2023},
  month = nov,
  journal = {Advances in Water Resources},
  volume = {181},
  pages = {104564},
  issn = {03091708},
  doi = {10.1016/j.advwatres.2023.104564},
  url = {https://linkinghub.elsevier.com/retrieve/pii/S0309170823001987},
  urldate = {2023-10-30},
  langid = {english}
}

@article{LiuA2024_PINN_Wells_Reservoir,
  title = {A Novel Reservoir Simulation Model Based on Physics Informed Neural Networks},
  author = {Liu, Aodi and Li, Jing and Bi, Jianfei and Chen, Zhangxing and Wang, Yan and Lu, Chunhao and Jin, Yan and Lin, Botao},
  year = {2024},
  month = nov,
  journal = {Physics of Fluids},
  volume = {36},
  number = {11},
  pages = {116617},
  issn = {1070-6631, 1089-7666},
  doi = {10.1063/5.0239376},
  url = {https://pubs.aip.org/pof/article/36/11/116617/3322481/A-novel-reservoir-simulation-model-based-on},
  urldate = {2025-03-31},
  langid = {english}
}

@article{LiuDC1989_LBFGS,
  title = {On the Limited Memory {{BFGS}} Method for Large Scale Optimization},
  author = {Liu, Dong C. and Nocedal, Jorge},
  year = {1989},
  month = aug,
  journal = {Mathematical Programming},
  volume = {45},
  number = {1-3},
  pages = {503--528},
  issn = {0025-5610, 1436-4646},
  doi = {10.1007/BF01589116},
  url = {http://link.springer.com/10.1007/BF01589116},
  urldate = {2025-04-18},
  copyright = {http://www.springer.com/tdm},
  langid = {english}
}

@article{LuL2021_PINN_HardConstraints,
  title = {Physics-{{Informed Neural Networks}} with {{Hard Constraints}} for {{Inverse Design}}},
  author = {Lu, Lu and Pestourie, Rapha{\"e}l and Yao, Wenjie and Wang, Zhicheng and Verdugo, Francesc and Johnson, Steven G.},
  year = {2021},
  month = jan,
  journal = {SIAM Journal on Scientific Computing},
  volume = {43},
  number = {6},
  pages = {B1105-B1132},
  issn = {1064-8275, 1095-7197},
  doi = {10.1137/21M1397908},
  url = {https://epubs.siam.org/doi/10.1137/21M1397908},
  urldate = {2024-01-09},
  langid = {english}
}

@misc{matplotlib_v291,
  title = {Matplotlib: {{Visualization}} with {{Python}}},
  shorttitle = {Matplotlib},
  author = {The Matplotlib Development Team},
  year = {2024},
  month = jul,
  doi = {10.5281/ZENODO.12652732},
  url = {https://zenodo.org/doi/10.5281/zenodo.12652732},
  urldate = {2025-04-30},
  copyright = {Creative Commons Attribution 4.0 International},
  howpublished = {Zenodo}
}

@book{MinskyM1972_MLP,
  title = {Perceptrons: An Introduction to Computational Geometry},
  shorttitle = {Perceptrons},
  author = {Minsky, Marvin and Papert, Seymour},
  year = {1969},
  edition = {2. print. with corr},
  publisher = {The MIT Press},
  address = {Cambridge/Mass.},
  isbn = {978-0-262-63022-1 978-0-262-13043-1},
  langid = {english}
}

@misc{Naumov2022_OpenGeoSys,
  title = {{{OpenGeoSys}} [{{Software}}]},
  author = {Naumov, Dmitry and Bilke, Lars and Fischer, Thomas and Rink, Karsten and Wang, Wenqing and Watanabe, Norihiro and Lu, Renchao and Grunwald, Norbert and Zill, Florian and Buchwald, J{\"o}rg and Huang, Yonghui and Bathmann, Jasper and Chen, Chaofan and Chen, Shuang and Meng, Boyan and Shao, Haibing and Kern, Dominik and Yoshioka, Keita and Rodriguez, Jaime and Miao, Xingyuan and Parisio, Francesco and Silbermann, Christian and Thiedau, Jan and Walther, Marc and Kaiser, Sonja and Boog, Johannes and Zheng, Tianyuan and Meisel, Tobias and Ning, Zhang},
  year = {2022},
  month = apr,
  doi = {10.5281/ZENODO.7092676},
  url = {https://doi.org/10.5281/zenodo.7092676},
  urldate = {2023-09-29},
  copyright = {BSD 3-Clause "New" or "Revised" License, Open Access},
  howpublished = {Zenodo}
}

@misc{pandas_v222,
  title = {Pandas-Dev/Pandas: {{Pandas}}},
  shorttitle = {Pandas-Dev/Pandas},
  author = {{The pandas development team}},
  year = {2024},
  month = apr,
  doi = {10.5281/ZENODO.10957263},
  url = {https://zenodo.org/doi/10.5281/zenodo.10957263},
  urldate = {2025-04-30},
  copyright = {BSD 3-Clause "New" or "Revised" License},
  howpublished = {Zenodo}
}

@misc{pytorch_v231,
  title = {{{PyTorch}}: {{An Imperative Style}}, {{High-Performance Deep Learning Library}}},
  author = {{Adam Paszke} and {Sam Gross} and {Francisco Massa} and {Adam Lerer} and {James Bradbury} and {Gregory Chanan} and {Trevor Killeen} and {Zeming Lin} and {Natalia Gimelshein} and {Luca Antiga} and {Alban Desmaison} and {Andreas K{\"o}pf} and {Edward Yang} and {Zach DeVito} and {Martin Raison} and {Alykhan Tejani} and {Sasank Chilamkurthy} and {Benoit Steiner} and {Lu Fang} and {Junjie Bai} and {Soumith Chintala}},
  year = {2024},
  month = jun,
  url = {https://github.com/pytorch/pytorch/releases/tag/v2.3.1},
  howpublished = {Meta}
}

@article{Raissi2019_PINN_Original,
  title = {Physics-Informed Neural Networks: {{A}} Deep Learning Framework for Solving Forward and Inverse Problems Involving Nonlinear Partial Differential Equations},
  author = {Raissi, M and Perdikaris, P and Karniadakis, G E},
  year = {2019},
  journal = {Journal of Computational Physics},
  volume = {378},
  pages = {686--707},
  publisher = {Elsevier Inc.},
  issn = {10902716},
  doi = {10.1016/j.jcp.2018.10.045},
  url = {https://doi.org/10.1016/j.jcp.2018.10.045}
}

@article{RoyP2024_ExactEnforcementTemporal,
  title = {Exact Enforcement of Temporal Continuity in Sequential Physics-Informed Neural Networks},
  author = {Roy, Pratanu and Castonguay, Stephen T.},
  year = {2024},
  month = oct,
  journal = {Computer Methods in Applied Mechanics and Engineering},
  volume = {430},
  pages = {117197},
  issn = {00457825},
  doi = {10.1016/j.cma.2024.117197},
  url = {https://linkinghub.elsevier.com/retrieve/pii/S0045782524004535},
  urldate = {2024-07-23},
  langid = {english}
}

@article{Rumelhart1986_MLP_Backrpopagation,
  title = {Learning Representations by Back-Propagating Errors},
  author = {Rumelhart, David E. and Hinton, Geoffrey E. and Williams, Ronald J.},
  year = {1986},
  month = oct,
  journal = {Nature},
  volume = {323},
  number = {6088},
  pages = {533--536},
  issn = {0028-0836, 1476-4687},
  doi = {10.1038/323533a0},
  url = {https://www.nature.com/articles/323533a0},
  urldate = {2025-04-08},
  copyright = {http://www.springer.com/tdm},
  langid = {english}
}

@article{Sarma2024_IPINNs_DomainDecomp,
  title = {Interface {{PINNs}} ({{I-PINNs}}): {{A}} Physics-Informed Neural Networks Framework for Interface Problems},
  shorttitle = {Interface {{PINNs}} ({{I-PINNs}})},
  author = {Sarma, Antareep Kumar and Roy, Sumanta and Annavarapu, Chandrasekhar and Roy, Pratanu and Jagannathan, Shriram},
  year = {2024},
  month = sep,
  journal = {Computer Methods in Applied Mechanics and Engineering},
  volume = {429},
  pages = {117135},
  issn = {00457825},
  doi = {10.1016/j.cma.2024.117135},
  url = {https://linkinghub.elsevier.com/retrieve/pii/S0045782524003918},
  urldate = {2025-05-14},
  langid = {english}
}

@article{ShenC2023_Differentiable_ML,
  title = {Differentiable Modelling to Unify Machine Learning and Physical Models for Geosciences},
  author = {Shen, Chaopeng and Appling, Alison P. and Gentine, Pierre and Bandai, Toshiyuki and Gupta, Hoshin and Tartakovsky, Alexandre and {Baity-Jesi}, Marco and Fenicia, Fabrizio and Kifer, Daniel and Li, Li and Liu, Xiaofeng and Ren, Wei and Zheng, Yi and Harman, Ciaran J. and Clark, Martyn and Farthing, Matthew and Feng, Dapeng and Kumar, Praveen and Aboelyazeed, Doaa and Rahmani, Farshid and Song, Yalan and Beck, Hylke E. and Bindas, Tadd and Dwivedi, Dipankar and Fang, Kuai and H{\"o}ge, Marvin and Rackauckas, Chris and Mohanty, Binayak and Roy, Tirthankar and Xu, Chonggang and Lawson, Kathryn},
  year = {2023},
  month = aug,
  journal = {Nature Reviews Earth \& Environment},
  volume = {4},
  number = {8},
  pages = {552--567},
  publisher = {Nature Publishing Group},
  issn = {2662-138X},
  doi = {10.1038/s43017-023-00450-9},
  url = {https://www.nature.com/articles/s43017-023-00450-9},
  urldate = {2025-03-27},
  copyright = {2023 Springer Nature Limited},
  langid = {english}
}

@article{Shukla2021_PINN_Domain_decomp,
  title = {Parallel Physics-Informed Neural Networks via Domain Decomposition},
  author = {Shukla, Khemraj and Jagtap, Ameya D. and Karniadakis, George Em},
  year = {2021},
  month = dec,
  journal = {Journal of Computational Physics},
  volume = {447},
  pages = {110683},
  issn = {00219991},
  doi = {10.1016/j.jcp.2021.110683},
  url = {https://linkinghub.elsevier.com/retrieve/pii/S0021999121005787},
  urldate = {2022-08-26},
  langid = {english}
}

@techreport{Slotte2017_LectureNotesWelltesting,
  title = {Lecture Notes in Well-Testing},
  author = {Slotte, Per Arne and Berg, Carl Fredrik},
  year = {2017},
  address = {Trondheim},
  institution = {NTNU}
}

@article{Soriano2021_AssessmentGroundwaterWell,
  title = {Assessment of Groundwater Well Vulnerability to Contamination through Physics-Informed Machine Learning},
  author = {Soriano, Mario A and Siegel, Helen G and Johnson, Nicholaus P and Gutchess, Kristina M and Xiong, Boya and Li, Yunpo and Clark, Cassandra J and Plata, Desiree L and Deziel, Nicole C and Saiers, James E},
  year = {2021},
  month = aug,
  journal = {Environmental Research Letters},
  volume = {16},
  number = {8},
  pages = {084013},
  issn = {1748-9326},
  doi = {10.1088/1748-9326/ac10e0},
  url = {https://iopscience.iop.org/article/10.1088/1748-9326/ac10e0},
  urldate = {2024-07-09}
}

@article{Sukumar2021_PINN_Hard_Constraints,
  title = {Exact Imposition of Boundary Conditions with Distance Functions in Physics-Informed Deep Neural Networks},
  author = {Sukumar, N. and Srivastava, Ankit},
  year = {2022},
  journal = {Computer Methods in Applied Mechanics and Engineering},
  volume = {389},
  pages = {114333},
  issn = {0045-7825},
  doi = {10.1016/j.cma.2021.114333},
  url = {https://www.sciencedirect.com/science/article/pii/S0045782521006186}
}

@article{SunJ2022_data_driven_ML_Groundwater,
  title = {Data-Driven Models for Accurate Groundwater Level Prediction and Their Practical Significance in Groundwater Management},
  author = {Sun, Jianchong and Hu, Litang and Li, Dandan and Sun, Kangning and Yang, Zhengqiu},
  year = {2022},
  month = may,
  journal = {Journal of Hydrology},
  volume = {608},
  pages = {127630},
  issn = {00221694},
  doi = {10.1016/j.jhydrol.2022.127630},
  url = {https://linkinghub.elsevier.com/retrieve/pii/S0022169422002050},
  urldate = {2023-11-01},
  langid = {english}
}

@article{Tartakovsky2020_PINN,
  title = {Physics-{{Informed Deep Neural Networks}} for {{Learning Parameters}} and {{Constitutive Relationships}} in {{Subsurface Flow Problems}}},
  author = {Tartakovsky, A M and Marrero, C Ortiz and Perdikaris, Paris and Tartakovsky, G D and Barajas-Solano, D},
  year = {2020},
  journal = {Water Resources Research},
  volume = {56},
  number = {5},
  issn = {0043-1397},
  doi = {10.1029/2019WR026731},
  url = {q}
}

@article{Taufik2024_Hard_constraints,
  title = {Stable Neural Network-Based Traveltime Tomography Using Hard-Constrained Measurements},
  author = {Taufik, Mohammad H. and Alkhalifah, Tariq and Waheed, Umair Bin},
  year = {2024},
  month = nov,
  journal = {GEOPHYSICS},
  volume = {89},
  number = {6},
  pages = {U87-U99},
  issn = {0016-8033, 1942-2156},
  doi = {10.1190/geo2024-0040.1},
  url = {https://library.seg.org/doi/10.1190/geo2024-0040.1},
  urldate = {2025-05-19},
  langid = {english}
}

@inproceedings{Teng2022_GreenFunction_ML,
  title = {Learning Green's Functions of Linear Reaction-Diffusion Equations with Application to Fast Numerical Solver},
  booktitle = {Proceedings of Mathematical and Scientific Machine Learning},
  author = {Teng, Yuankai and Zhang, Xiaoping and Wang, Zhu and Ju, Lili},
  editor = {Dong, Bin and Li, Qianxiao and Wang, Lei and Xu, Zhi-Qin John},
  year = {2022},
  series = {Proceedings of Machine Learning Research},
  volume = {190},
  pages = {1--16},
  publisher = {PMLR},
  url = {https://proceedings.mlr.press/v190/teng22a.html}
}

@article{Vaezi2024_Bedretto_Stimulation_Model,
  title = {Numerical Modeling of Hydraulic Stimulation of Fractured Crystalline Rock at the Bedretto Underground Laboratory for Geosciences and Geoenergies},
  author = {Vaezi, Iman and Alcolea, Andr{\'e}s and Meier, Peter and Parisio, Francesco and Carrera, Jesus and Vilarrasa, V{\'i}ctor},
  year = {2024},
  journal = {International Journal of Rock Mechanics and Mining Sciences},
  langid = {english}
}

@book{Wang2000,
  title = {Theory of Linear Poroelasticity with Applications to Geomechanics and Hydrogeology},
  author = {Wang, Herbert F},
  year = {2000},
  publisher = {Princeton Univ. Press},
  address = {Princeton},
  isbn = {0-691-03746-9}
}

@article{WangN2020_PINN_subsurface_flow,
  title = {Deep Learning of Subsurface Flow via Theory-Guided Neural Network},
  author = {Wang, Nanzhe and Zhang, Dongxiao and Chang, Haibin and Li, Heng},
  year = {2020},
  journal = {Journal of Hydrology},
  volume = {584},
  number = {January},
  pages = {124700},
  publisher = {Elsevier},
  issn = {0022-1694},
  url = {https://doi.org/10.1016/j.jhydrol.2020.124700}
}

@article{WangN2024_PICD_flow_inversion,
  title = {Physics-{{Informed Convolutional Decoder}} ({{PICD}}): {{A Novel Approach}} for {{Direct Inversion}} of {{Heterogeneous Subsurface Flow}}},
  shorttitle = {Physics-{{Informed Convolutional Decoder}} ({{PICD}})},
  author = {Wang, Nanzhe and Kong, Xiang-Zhao and Zhang, Dongxiao},
  year = {2024},
  month = jul,
  journal = {Geophysical Research Letters},
  volume = {51},
  number = {13},
  pages = {e2024GL108163},
  issn = {0094-8276, 1944-8007},
  doi = {10.1029/2024GL108163},
  url = {https://agupubs.onlinelibrary.wiley.com/doi/10.1029/2024GL108163},
  urldate = {2024-12-05},
  langid = {english}
}

@article{WangS2023_ExpertGuidePINN,
  title = {An {{Expert}}'s {{Guide}} to {{Training Physics-informed Neural Networks}}},
  author = {Wang, Sifan and Sankaran, Shyam and Wang, Hanwen and Perdikaris, Paris},
  year = {2023},
  journal = {Preprint},
  doi = {10.48550/ARXIV.2308.08468},
  url = {https://arxiv.org/abs/2308.08468},
  urldate = {2023-12-09},
  copyright = {Creative Commons Attribution Non Commercial Share Alike 4.0 International}
}

@article{WangY2024_MultistageNeuralNetworks,
  title = {Multi-Stage Neural Networks: {{Function}} Approximator of Machine Precision},
  shorttitle = {Multi-Stage Neural Networks},
  author = {Wang, Yongji and Lai, Ching-Yao},
  year = {2024},
  month = may,
  journal = {Journal of Computational Physics},
  volume = {504},
  pages = {112865},
  issn = {00219991},
  doi = {10.1016/j.jcp.2024.112865},
  url = {https://linkinghub.elsevier.com/retrieve/pii/S0021999124001141},
  urldate = {2024-11-21},
  langid = {english}
}

@incollection{WatsonAT1994_ParameterSystemIdentification,
  title = {{Parameter and System Identification for Fluid Flow in Underground Reservoirs}},
  booktitle = {{Proceedings of the Conference Inverse Problems and Optimal Design in Industry}},
  author = {Watson, A. T. and Wade, J. G. and Ewing, R. E.},
  editor = {Arkeryd, Leif and Engl, Heinz and Fasano, Antonio and Mattheij, Robert M. M. and Neittaanm{\"a}ki, Pekka and Neunzert, Helmut and Engl, Heinz W. and McLaughlin, Joyce},
  year = {1994},
  pages = {81--108},
  publisher = {Vieweg+Teubner Verlag},
  address = {Wiesbaden},
  doi = {10.1007/978-3-322-96658-2_5},
  url = {https://link.springer.com/10.1007/978-3-322-96658-2_5},
  urldate = {2024-08-22},
  isbn = {978-3-322-96659-9 978-3-322-96658-2},
  langid = {ngerman}
}

@article{YanX2024_PINN_fractured_reservoir,
  title = {Physics-{{Informed Neural Network Simulation}} of {{Two-Phase Flow}} in {{Heterogeneous}} and {{Fractured Porous Media}}},
  author = {Yan, Xia and Lin, Jingqi and Wang, Sheng and Zhang, Zhao and Liu, Piyang and Sun, Shuyu and Yao, Jun and Zhang, Kai},
  year = {2024},
  month = may,
  journal = {Advances in Water Resources},
  pages = {104731},
  issn = {03091708},
  doi = {10.1016/j.advwatres.2024.104731},
  url = {https://linkinghub.elsevier.com/retrieve/pii/S0309170824001180},
  urldate = {2024-05-27},
  langid = {english}
}

@article{Zhang2022_GWPINN,
  title = {{{GW-PINN}}: {{A}} Deep Learning Algorithm for Solving Groundwater Flow Equations},
  shorttitle = {{{GW-PINN}}},
  author = {Zhang, Xiaoping and Zhu, Yan and Wang, Jing and Ju, Lili and Qian, Yingzhi and Ye, Ming and Yang, Jinzhong},
  year = {2022},
  month = jul,
  journal = {Advances in Water Resources},
  volume = {165},
  pages = {104243},
  issn = {03091708},
  doi = {10.1016/j.advwatres.2022.104243},
  url = {https://linkinghub.elsevier.com/retrieve/pii/S0309170822001142},
  urldate = {2024-07-09},
  langid = {english}
}

\end{document}